\documentclass[sigconf]{acmart}
\newcommand{\X}{{\mathbf {X}}}
\newcommand{\XJL}{{\mathbf {Y}}}
\newcommand{\IE}{\mathbb{E}}

\newcommand{\IP}{\mathbb{P}}
\newcommand{\randeigen}{\textsc{RandEigen}}
\usepackage{algorithm}
\usepackage{ algorithmicx}
\usepackage{algcompatible}
\usepackage{mdframed}
\usepackage{svg}
\usepackage{subfig}
\definecolor{mygray}{RGB}{211,211,211} 
\newtheorem*{theorem*}{Theorem}
\newtheorem{remark}{Remark}
\newtheorem{assumption}{Assumption}
\AtBeginDocument{%
  }
\usepackage{multirow}

\usepackage[compact]{titlesec}         
\titlespacing{\section}{1pt}{1pt}{1pt} 
\AtBeginDocument{
  \setlength\abovedisplayskip{1pt}
  \setlength\belowdisplayskip{1pt}}

\copyrightyear{2025}
\acmYear{2025}
\setcopyright{acmlicensed}
\acmConference[CCS '25] {Proceedings of the 2025 ACM SIGSAC Conference on Computer and Communications Security}{ October 13--17, 2025}{Taipei, Taiwan.}
\acmBooktitle{Proceedings of the 2025 ACM SIGSAC Conference on Computer and Communications Security (CCS '25), October 13--17, 2025, Taipei, Taiwan}
\acmISBN{979-8-4007-1525-9/2025/10}
\acmDOI{10.1145/3719027.3744851}
\usepackage{xcolor}

\usepackage{subfiles} 
\author{De Zhang Lee}
\affiliation{%
  \institution{National University of Singapore}
  \country{Singapore}
  }
\email{dezhanglee@comp.nus.edu.sg}

\author{Aashish Kolluri}
\affiliation{%
  \institution{National University of Singapore}
  \country{Singapore}
  }
\email{aashishk@u.nus.edu}

\author{Prateek Saxena}
\affiliation{%
  \institution{National University of Singapore}
  \country{Singapore}
  }
\email{prateeks@comp.nus.edu.sg}

\author{Ee-Chien Chang}
\affiliation{%
  \institution{National University of Singapore}
  \country{Singapore}
  }
\email{changec@comp.nus.edu.sg}
\begin{document}

\title{A Practical and Secure Byzantine Robust Aggregator}









\begin{abstract}

In machine learning security, one is often faced with the problem of removing outliers from a given set of high-dimensional vectors when computing their average. For example, many variants of data poisoning attacks produce gradient vectors during training that are outliers in the distribution of clean gradients, which bias the computed average used to derive the ML model. Filtering them out before averaging serves as a generic defense strategy. Byzantine robust aggregation is an algorithmic primitive which computes a robust average of vectors, in the presence of an $\epsilon$ fraction of vectors which may have been arbitrarily and adaptively corrupted, such that the resulting bias in the final average is provably bounded. 

In this paper, we give the first robust aggregator that runs in quasi-linear time in the size of input vectors and provably has near-optimal bias bounds. Our algorithm also does not assume any knowledge of the distribution of clean vectors, nor does it require pre-computing any filtering thresholds from it. This makes it practical to use directly in standard neural network training procedures. 
We empirically confirm its expected runtime efficiency and its effectiveness in nullifying 10 different ML poisoning attacks.
    
\end{abstract}

\maketitle

\section{Introduction} \label{section:intro}


Machine learning (ML) models are trained on massive datasets gathered from the open web. A prominent threat to such training is data poisoning attacks~\cite{DBLP:conf/icml/BiggioNL12}. In such attacks a small fraction of the training data samples, or the gradient vectors derived from them, are adversarially corrupted such that the final model deviates from the original. The objective of effecting such deviation varies across attacks. Backdoor poisoning attacks aim for ML models to recognize secret patterns  that cause errant outputs at inference time. Untargeted poisoning attacks cause mis-classification degrading accuracy during training, while targeted poisoning attacks cause misclassification selectively. Recent works have argued that web-scale poisoning is practical with very low cost \cite{DBLP:conf/sp/CarliniJCPPATTT24}. 

We want {\em generic} and {\em principled} defenses that can tackle a large variety of poisoning attacks. Very few ideas for such defenses are presently known. One promising approach is based on {\em Byzantine robust aggregation}. A robust aggregator takes as input a set of vectors, a small fraction $\epsilon$ of which may have been arbitrarily corrupted to deviate from the rest, and computes a robust mean for this set \cite{DBLP:journals/corr/abs-1911-05911}. The robust mean is guaranteed to not deviate much from the mean of the uncorrupted vectors in the input set, i.e., the norm (magnitude) of the deviation caused by the corrupted vectors is tightly bounded. This deviation is also called the {\em bias}.

If we have such a robust aggregator, it can be used to average the gradient vectors during standard training such as with stochastic gradient descent or with other methods that use averaging. This approach is compatible both with centralized training, where all data is gathered at a central service for training, as well as the federated learning setup wherein multiple distributed workers send locally computed gradient vectors to an aggregation server.
The strategy of norm-bounding the deviation of the average of gradient vectors offers a generic and principled way to beat many attacks---several prior works \cite{DBLP:journals/corr/abs-1911-05911, zhu2023byzantine,shen2016auror,dnc,fang2020local, DBLP:journals/pami/GoldblumTXCSSML23} have made this observation.

Designing a Byzantine robust aggregator is a central focus of this paper. The main challenge is that of aggregating over high-dimensional vectors. Gradient vectors arising in training of modern neural network models have $d$ dimensions, typically higher than $2^{30}$. Robust aggregators that run in $\Omega(d^2)$ are computationally too expensive to use in practice. Similarly, aggregators with large bounds on the bias have weak security. For example, there are linear-time  aggregators that can upper bound the bias to a factor $O(\sqrt{d})$ of the variance of the uncorrupted vectors. A recent attack shows that such vacuously large bounds can permit real poisoning attacks in the federated learning setup, causing extensive loss in model performance during training \cite{DBLP:conf/sp/ChoudharyKS24}.
Our goal is to devise techniques that have quasi-linear running time $\tilde{O}(n\cdot d)$, where $n$ is the number of vectors and $d$ is the number of dimensions, while having dimension-independent bound on the maximum possible bias. 

\paragraph{Practical Robust Aggregation.}
In the statistics literature, the concept of ``robustness'' was formally 
introduced in 1953 \cite{box_robust}. 
However, techniques such as dimension-wise mean aggregation to mitigate the 
influence of outliers in experiments have been in use for over a 
century---early examples include Newcomb's work in 1886 \cite{newcomb1886generalized}. 
Tukey's seminal work on robust statistics \cite{tukey1960survey} 
 observed that such techniques yield a sub-optimal 
bias upper bound, and 
generalized the notion of the statistical median to 
higher dimensions through the Tukey median which achieves the information-theoretic
optimal bias. Despite its theoretical significance, 
the exact computation or approximation of the Tukey median is NP-Hard 
\cite{DBLP:conf/colt/Hopkins019}, making it impractical to be employed
as a Byzantine robust aggregator. 

Table~\ref{table:priorwk} summarizes the state-of-the-art solutions. The bias characterizes the theoretical security guarantee offered. It is an upper bound, a multiplicative factor of the natural variance of the uncorrupted (clean) samples in the set of vectors given as input. Weak robust aggregators have dimension-dependent bounds on the bias which give a weak theoretical security. Strong aggregators limit the adversarial bias in the computed mean to near optimal, i.e., independent of $d$. The best known strong robust aggregators run in $\Omega(nd^2)$ which are prohibitively slow in high-dimensional setups.

\begin{table}[tbp]
    \centering
    \caption{Maximum bias and computational complexity of several weak and strong robust aggregators.
    $||\Sigma||_2$ refers to the maximum variance of the uncorrupted sample
    (dominant eigenvalue of the covariance matrix) over the given $d$-dimensional $n$ vector samples. $\tilde{O}( \cdot)$ ignores the constant and logarithmic factors in the computation complexity.} 
    \label{table:priorwk}
    \renewcommand{\arraystretch}{1.5}
    \scalebox{0.8}{
    \begin{tabular}{|c|c|c|}
        \hline
        \textbf{Algorithm} & \textbf{Max. Bias} & \textbf{Comp. Complexity} 
        \\
        \hline
        Weak Robust Aggregators& &
        \\
        \cline{1-1} 
        \label{row:suboptimal_start}Median ~\cite{yin2018byzantine} & & \\
       Trimmed Mean ~\cite{yin2018byzantine}& $\tilde{O}(\sqrt{d}) \cdot \sqrt{||\Sigma||_2}$ & $\tilde{O}(nd)$
        \\
        \label{row:suboptimal_end} Geometric Median ~\cite{pillutla2022robust, blanchard2017machine, chen2020distributed} & & 
        \\
        \hline
        Strong Robust Aggregators& &
        \\
        \cline{1-1} 
    \label{row:optimal_start}Filtering~\cite{diakonikolas2017being}&  & $\tilde{O}(\epsilon n d^3)$ \\
        No-Regret~\cite{hopkins2020robust} &  & $\tilde{O}((n + d^3)\cdot d)$ 
        \\
        \label{row:optimal_end}SoS ~\cite{kothari2017outlierrobust} & $\tilde{O}(1)\cdot \sqrt{||\Sigma||_2}$ & poly$(n, d)$ 
        \\
        Tukey Median~\cite{tukey1960survey} & & NP-Hard in $d$ 
        \\
        RandEigen &  & $\tilde{O}(nd)$
        \\
        \hline
    \end{tabular}
    }
\end{table}

The most computationally efficient among strong aggregators are those that employ iterative filtering. In this approach, outlier vectors are iteratively filtered out, or assigned lower weights, before computing their final average.  In each iterateion, iterative filtering works by computing the dominant eigenvector\footnote{The dominant eigenvector is the of eigenvector with largest eigenvalue, which is also the the one among all $d$-dimensional unit vectors that maximizes the variance across projections of the given input vectors onto the unit vector.}
of the covariance matrix of the given vectors. Vectors which are far away from the mean of the projections along the dominant eigenvector are filtered out, if they exceed a pre-computed threshold. The process repeats for $n \epsilon$ steps and the average of the vectors that remain is the result.

There are $2$ practical limitations of iterative filtering.
First, the pre-computed threshold is a constant that is explicitly computed from the variance of the uncorrupted vectors. In order to use iterative filtering, one has to determine the constant threshold apriori, for example, by observing the variance of the gradient distributions under clean training. This is a practical hinderance since the trainer may not have access to clean samples to begin with, or have resources to train with them. The second severe limitation is that the running time of this procedure is $\Omega(nd^2)$, owing to the computational cost of computing the dominant eigenvector.

\paragraph{Our Contribution.}
We give \textsc{RandEigen}, the first iterative filtering algorithm with $\tilde{O}(nd)$ running time and dimension-independent bias bound. The running time thus is quasi-linear in the input size and has near-optimal bias. It also requires no knowledge of the properties of the clean sample distribution. 
We introduce $2$ key improvements into the original iterative filtering algorithm. The first is a fast randomized procedure to compute the dominant eigenvector. The second is to replace the pre-computed threshold with a convergence check. The latter \emph{removes the need for pre-computing thresholds}. It also mitigates prior attacks that target deterministic filtering thresholds.  We prove that \textsc{RandEigen}, which is a randomized procedure, gives correct robust aggregates on expectation.

\paragraph{Experimental Evaluation.}
We confirm the practicality and effectiveness of \textsc{RandEigen} experimentally as well. We focus our evaluation for the federated learning setup wherein the attacker directly corrupts gradient vectors during training. We evaluate on commonly used image classification tasks as well as on some potent attacks on language models.
These include targeted, untargeted, and backdoor attacks that are not-adaptive, i.e., they are not designed specifically to beat strong aggregators evaluated in prior works. We also evaluate the recent HiDRA attack which is an adaptive attack that aims to create optimal bias against deterministic aggregators specifically.
Our findings are consistent with our theoretical expectations---all of these attacks are generically mitigated by \textsc{RandEigen}. In terms of computational overhead, {\sc RandEigen} has a running time that is about $15\times$ 
faster than the previously proposed strong robust iterative filtering even in low dimensional setting of $d=1000$. The gap widens as dimensions increase, for example, \textsc{RandEigen} is more than $300\times$ faster when $d > 8000$.
We therefore claim that \textsc{RandEigen} is feasible to use in standard training. We further report on experiments against attacks in the centralized setting, showing strong effectiveness (see Section \ref{subsection:eval_centralized}). This demonstrates that \randeigen~works well in both federated and centralized setups. The drop in model accuracy is small ($\leq 2\%$)  when \randeigen~is used and there are no attacks.

\paragraph{Caveat.}
Our contribution is on making bias-bounding defenses practical to use. This class of defenses works by limiting magnitude of bias in gradient averaging, which has been empirically observed in many poisoning attacks. That said, this property by itself does not imply a complete defense against all possible poisoning attacks. For example, it is possible that there exist attacks that do not bias the model updates much in each training step much and yet achieve a malicious objective overall. A systematic investigation of such attacks is promising future work but beyond the scope of this work.

\section{Motivation \& Problem Definition}
\label{sec:motivation}

Our pursuit for designing efficient robust aggregators stems from designing a generic defense for data poisoning attacks.

\subsection{Poisoning Attacks}

Poisoning attacks can be broadly categorized into three types \cite{DBLP:journals/ieeesp/JereFK21}: Backdoor, targeted, and untargeted attacks. 
Backdoor and targeted attacks aim to induce specific misclassifications when certain trigger features are present in the input, 
while untargeted attacks seek to degrade the overall accuracy of the model. 
\begin{figure*}[!htbp]
    \centering
    \includegraphics[width=0.247\linewidth]{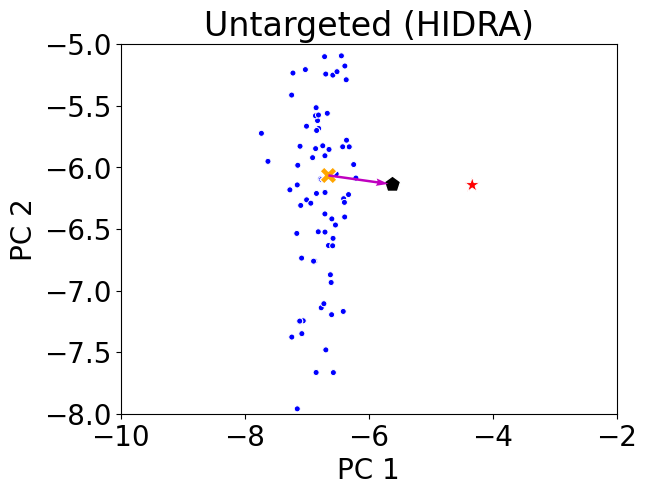}
    \includegraphics[width=0.247\linewidth]{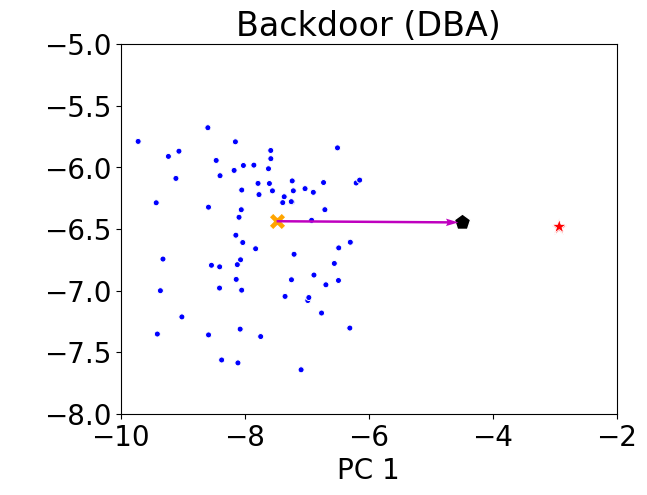}
    \includegraphics[width=0.247\linewidth]{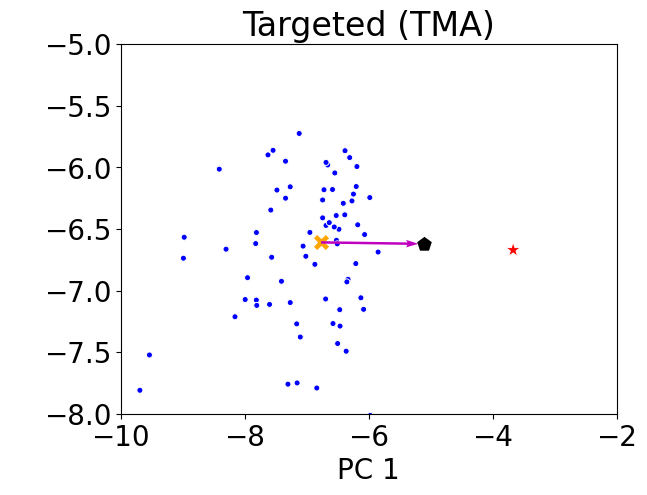}
    \includegraphics[width=0.247\linewidth]{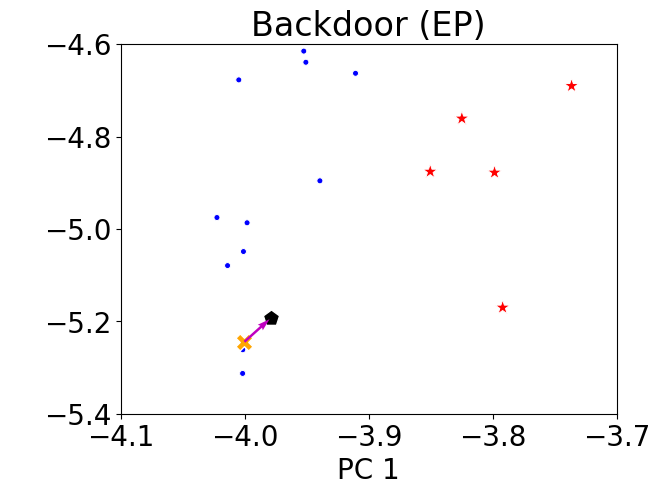} \\
    \includegraphics[width=0.7\linewidth]{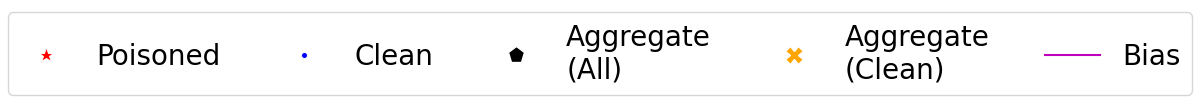}
    \caption{Illustrations of the gradients gradients sampled at random steps of SGD under three different data poisoning attacks, 
    on image classification (HIDRA \cite{DBLP:conf/sp/ChoudharyKS24},
    DBA \cite{xie2019dba}, TMA \cite{fang2020local}) and language 
    (EP \cite{yang-etal-2021-careful}) models. 
    PC 1 and PC 2 represent the projections (on a log scale)
    of the gradients onto the eigenvectors corresponding to the largest and second-largest eigenvalues, respectively.
    }
    \label{fig:intro_comparison}
\end{figure*}

\begin{algorithm}[H]
\caption{Federated Averaging (FedAvg) \cite{DBLP:conf/aistats/McMahanMRHA17}
using SGD.}
\label{alg:fedavg}
\begin{algorithmic}[1]
\STATE \textbf{Input:} Number of training steps $T$, clients per step $K$, learning rate $\eta$, initial model parameters $w_0$, robust aggregation function \randeigen

\FOR{step $t = 0, 1, \dots, T-1$}
    \STATE Server sends the current global model $w_t$ to all $k$ clients
    \FOR{each client $k \in \mathcal{S}_t$ \textbf{in parallel}}
        \STATE Client $k$ trains $w_t$ locally, compute
        gradient of loss function $\nabla \mathcal{L}_k(w_t)$
        \STATE Send gradients $\nabla \mathcal{L}_k(w_t)$ back to the server.
    \ENDFOR
    \STATE  $\nabla \mathcal{L}_{t} = \randeigen(\nabla \mathcal{L}_1(w_t), \dots, \nabla \mathcal{L}_k(w_t))$ \label{alg:fedavg_line_agg}
    \STATE Server computes new parameters $w_{t+1} = w_t - \eta \nabla \mathcal{L}_{t}$
\ENDFOR

\STATE \textbf{Output:} Final global model with parameters $w_T$.
\end{algorithmic}
\end{algorithm}

Prior works (e.g. \cite{DBLP:journals/corr/abs-1911-05911,shen2016auror,li20233dfed,dnc,fang2020local, DBLP:journals/pami/GoldblumTXCSSML23}) have observed a common property of many such attacks: The gradients of corrupted samples are often much larger in magnitude than those of clean samples. Figure
\ref{fig:intro_comparison} plots the gradient vectors obtained at a randomly chosen step during training with stochastic gradient descent of recent representative attacks, covering all of the $3$ kinds. It can be readily seen that the gradients of the poisoned data samples are outliers, significantly larger than those of the gradients of clean samples. The final gradient average computed is thus biased during gradient descent and causes the next ML model update to be skewed as well.
It is natural, therefore, to consider removing  gradient outliers as a generic defense against poisoning attacks.

Algorithm \ref{alg:fedavg} shows a slight modification of the stochastic gradient descent algorithm (SGD) to include the envisioned defense.
Line \ref{alg:fedavg_line_agg} computes an average of the gradients, much like the original, except that this average is a robust mean as defined in Section~\ref{section:proposed_randeigen}. It is responsible for filtering out the outlier gradients in the computation of the average, thereby limiting the impact of the malicious or poisoned gradients on the model update. Throughout this paper we will consider SGD as the training procedure, though the main idea can be incorporated into other popular training methods such as Adam~\cite{adam} which also use gradient averaging, see Appendix \ref{appendix:adam}.


\paragraph{Concrete Threat Model.}
Two different setups that are considered in the context of poisoning attacks: centralized and federated learning.
For centralized, we are mostly interested in the data poisoning attacks wherein data samples are poisoned. Recent work has shown that poisoning about $1-5\%$ of the samples
is fairly practical~\cite{DBLP:conf/sp/CarliniJCPPATTT24}.
In federated learning, a model is collaboratively trained across multiple participants,
some of whom may be malicious. 
Several studies (e.g., \cite{tolpegin2020data, DBLP:conf/sp/ChoudharyKS24}) have shown that data poisoning attacks can be effectively carried out with as few as 5\% of clients being malicious.

Both setups employ averaging of gradient vectors computed from data samples.
For example, consider the standard stochastic gradient descent method for training neural networks. In the centralized setting, the gradient is computed and averaged
across batches. 
In federated learning, the same algorithm can be used but with gradients being computed locally at each worker machine. At the end of each training step, 
the gradients across all workers are averaged and used to update the central model.
If we assume that some of the workers are potentially malicious, they may 
deliberately send incorrect gradients to the centralized server with the 
intent of biasing the central ML model.
Byzantine attacks against FL can be viewed as a data poisoning attack, where
Byzantine clients construct poisoned updates according to their goals. 

In summary, we are interested in designing a defense that is flexible enough to be used in standard training procedures across both centralized and federated setup in practice. For concreteness, however, we will evaluate in the federated learning setup that gives the adversary the most power: The adversary directly controls the gradient vectors, knows the training procedure, and the distribution of the clean samples. It knows everything that the defender (trainer) knows except the randomness used by the defender, but is computationally bounded to run in polynomial time in the size of the inputs. The adversary is allowed to be fully adaptive. 

The defender is the party that runs the aggregation step (e.g. server in Algorithm~\ref{alg:fedavg}) and is honest. All attacks that bias the gradients significantly are considered in scope for the defense. Attacks violating other properties without violating the bias bounds are outside the scope of our defense.
The attacker knows everything the defender does, and the coin flips of the aggregator are unbiased.

The defense goal 
is to mitigate the impact of Byzantine attacks by
utilizing a robust aggregator that minimizes the bias introduced by 
malicious gradients. 
Depending on the type of attack, the defender's success is measured in the standard way as follows:
\begin{itemize}
    \item Targeted/Backdoor attacks: Model accuracy on non-targeted data and the attack success rate (ASR). ASR is the proportion of samples containing the targeted feature or backdoor trigger that are classified as the attacker intended class.
    \item Untargeted attacks: Overall model accuracy.
\end{itemize}

\subsection{Byzantine Robust Aggregation}

Byzantine Robust aggregation is the problem of computing a mean statistic for a given set of vectors, where up to an $\epsilon$-fraction of vectors may be corrupted using
some corruption strategy with the goal of introducing a large bias on the estimated 
aggregate.

Formally, consider a set $X$ of $n$ vectors in $\mathbb{R}^d$.
An $\epsilon$ fraction of $n$ vectors are {\em arbitrarily} corrupted, consistent with Huber's contamination model \cite{10.1214/aoms/1177699803}, and given as input to a Byzantine robust aggregator function $f:\mathbb{R}^{n\times d} \rightarrow \mathbb{R}^d$. Let $S \subseteq X$ be the uncorrupted vectors in $X$. The output of $f$ must satisfy the following property:

$$
|| f(X) - \frac{1}{|S|}\sum_{s \in S} s||_2 \leq \beta \cdot  \sqrt{||\Sigma_S||_2}
$$

The quantity on the left hand side of the inequality above is the {\em bias}, the magnitude of the difference between final mean $f(X)$ and the mean of the uncorrupted samples $S$.
The quantity on the right is the upper bound on the maximum possible bias under all adversarial strategies. The uncorrupted sample have some natural (benign) variance, which is given by $|| \Sigma_S ||_2$. It is the standard spectral norm of the sample covariance matrix computed from $S$.

$S$ is not known to the defender who runs $f$, otherwise it would trivially  eliminate the poisoned vectors in $X$ and ensure zero bias.

The key quantity for security is $\beta$, the multiplicative factor that bounds the maximum bias. We want $\beta$ to be as small as possible for the tightest defense. In the literature,
there are broadly two classes of robust aggregators: {\em weak} and {\em strong}. Strongly-bounded (or strong) robust aggregators
can ensure that $\beta$ is $\tilde{O}(1)$ whereas $\beta$ for weakly-bounded (or weak) aggregators is $\tilde{O}(\sqrt{d})$. The difference between strong and robust aggregators is that the latter admits bias that grows with $\sqrt{d}$. In modern neural network training, $d$ is typically higher than $2^{30}$, which means that the $\beta$ can be many orders of magnitude worse for weak aggregators if used as a defense. Our focus, therefore, is primarily on designing strong aggregators with $\tilde{O}(1)$ factor in the bias, which is statistically optimal.




\paragraph{Assumptions.}
We make the following assumptions, which are the same as in prior works~\cite{DBLP:journals/corr/abs-1911-05911}:
\begin{enumerate}
    \item \textbf{Fraction of corruption $\epsilon$}. We assume that $\epsilon < 1/12$ for our theoretical analysis. 
    However, our empirical evaluation demonstrates that the proposed Byzantine-robust aggregator remains effective even when $\epsilon = 0.20$.
    \item \textbf{Unknown distribution of clean samples}. 
    The most efficient strong aggregators from the literature assume that the benign variance $||\Sigma_S||_2$ of the clean samples $S$ is somehow computable by the defender. This can difficult to satisfy in practice, since the defender may not have access to the distribution of clean samples apriori. Even if they do, pre-computing this quantity for each step during training requires running the training with samples from the clean distribution first. In our work, we do not require computing the benign variance $||\Sigma_S||_2$ explicitly. We retain the same assumptions as prior work (e.g. \cite{DBLP:journals/corr/abs-1911-05911, DBLP:conf/icml/DiakonikolasKK019}), namely that the set of clean samples is unknown but has (1) finite mean and covariance, and (2) is $(5\epsilon, \delta)$-stable, stated formally in Section \ref{subsection:prelim_assumption}. Here, $\delta$ is within a constant factor of $\sqrt{||\Sigma_S||_2}$.
    Roughly speaking, (2) implies that the mean of any 
    subset of the clean
    samples (with size at least $1-5\epsilon$ fraction of the full set) deviates from
    the mean over all samples by at most $\delta$.
     
\end{enumerate}

The aggregator $f$ must have practical running time for high-dimensional inputs. We give the first strong robust aggregator that runs in $\tilde{O}(nd)$. The previous strong aggregators all run in time 
$\Omega(nd^2)$, to the best of our knowledge. Please see Table~\ref{table:priorwk}.

\section{Background: Iterative Filtering}

Our work builds upon the original 
iterative filtering algorithm proposed in \cite{DBLP:journals/corr/abs-1911-05911}, which is a strong robust aggregator running in polynomial-time.
It takes as input $X = S \cup T$ be an $\epsilon$-corrupted
set consisting of $n$ $d$-dimensional samples, 
for $\epsilon<1/12$, where $S$ and $T$ are the clean and poisoned
samples, respectively. Let $(\mu_S, \mu_T, \mu_X)$ and 
$(\Sigma_S, \Sigma_T,$ $\Sigma_X)$ denote
the sample mean and sample covariance matrix of $S, T,$ and $X$, respectively in the exposition below. 

\subsection{Baseline: Iterative Filtering Algorithm} \label{section:original_filtering}
The most efficient prior algorithm for strong aggregation is the \texttt{Randomized Filtering}
algorithm given in~\cite{diakonikolas2023algorithmic}, which we present in Algorithm \ref{alg:filter_original}. It is a randomized algorithm and it assumes a known estimate of the upper bound of $||\Sigma_S||_2$. In this iterative algorithm, points in $X$ projecting along the principal component $u$, i.e., along the dominant eigenvector of the sample covariance matrix of $X$.
%
Each point is removed from $X$ with a probability proportional to its projection along the dominant eigenvector---outliers will have larger projected components and are more likely to be filtered out. The process repeats until either (1) the number of iterations reaches $2n\epsilon$, or (2) the largest eigenvalue ($\lambda_{curr}$) of the sample covariance matrix of the set $X$
falls below some predetermined threshold $\Gamma$.
The value of $\Gamma = k ||\hat{\Sigma}_S||_2$ for $k \in [\sqrt{20}, 9]$, where $\hat{\Sigma}_S$ is the empirical
covariance matrix estimated from some other clean samples\cite{DBLP:conf/sp/ChoudharyKS24}.

\setlength{\textfloatsep}{0pt}
\begin{algorithm}[t]
\caption{Pseudo-code for the original \texttt{Randomized Filtering}.}
\label{alg:filter_original}
\begin{algorithmic}[1] 
\REQUIRE Fraction of corruption $\epsilon$, 
set of $\epsilon$-corrupted updates represented by $n \times d$ matrix $X = \{x_1, \dots, x_d\}$, $\Gamma=$ Upper bound of $||\Sigma_S||_2$
\ENSURE Robust aggregate $\mu = \{\mu_1, \dots, \mu_d\}$, where $\mu_i$ is the robust aggregate of the $i$-th dimension

\FOR{$j=1$ {\bfseries to} $ 2n \cdot \epsilon$} 
    \STATEx \quad $\triangleright$  Obtain dominant eigenvalue/vector
    \STATE $\lambda_{curr}, u := EIGENDECOMPOSITION(Cov(X))$ 
    \IF{$\lambda_{curr} < \Gamma$} \label{algo:stop_ori} 
         break \qquad \qquad $\triangleright ||\Sigma_X||_2 < \Gamma$  
    \ENDIF
    \STATE $\mu_x = \frac{1}{|X|} \sum_{x \in X} x$
    \STATE  $\mathcal{P} = \{p_1, \dots, p_n\}$, where $p_i = |\langle x_i, u \rangle -  \langle \mu_x, u  \rangle|$
    \FOR {$i= 1$ {\bfseries to} $ |X|$} 
        \STATE Remove $x_i$ from $X$ with probability $p_i / \max \mathcal{P}$
    \ENDFOR
\ENDFOR
\State \textbf{return} $\mu$ = Dimension-wise average of $X$
\end{algorithmic}
\end{algorithm}

The reason why Algorithm
\ref{alg:filter_original} works is as follows: 
To significantly bias $\mu_X$ away from $\mu_S$,
$\mu_T$ must be of some significant magnitude---call it $\delta$. Specifically,
the $n\epsilon$ corrupted vectors $T$ must pull together in some direction in order to create a bias larger than the benign variance of a set of $n\epsilon$ benign vectors. It can then be shown that corrupted vectors $T$ must induce a total increase in variance of at least $\frac{\delta^2}{\epsilon}$ along some direction.
From properties of principal component analysis, it is known that the dominant eigenvector of the covariance matrix of $X$ {\em is the projection direction that exhibits the maximum average variation among points in $X$}. So, if there exists any direction along which the projection of $\mu_T$ is larger than $\delta$, and consequently the corresponding variance along that direction is large (exceeds  $\frac{\delta^2}{\epsilon}$), then the variation of $X$ must be at least that large along the dominant eigenvector direction. Two properties follow from this fact. First, if the check on Line 3 passes, the largest eigenvalue being lesser than the threshold $\Gamma=k\cdot||\Sigma_S||_2$, there exists no direction along which the spread of the points remaining in $X$ exceeds the  $\Gamma$ (which is near-optimal as $\Gamma$ is a constant factor of the benign variance).
Second, the projected components of outlier points along the dominant eigenvector will be larger than those of inliers, causing them to be filtered out with proportionally higher probability on Line 8. In particular, 
the probability of the farthest outlier along the dominant eigenvector direction will be removed with probability $1$ on Line 8. Overall, the guarantee is that
with high probability that the final $X$ after filtering will have variance close to the benign variance $||\Sigma_S||_2$. This implies the computed mean of filtered $X$ is close to $\mu_S$. We refer readers to Chapter 2.4 in~\cite{diakonikolas2023algorithmic} for formal details.

A noteworthy point is that it is necessary to recompute the direction of the dominant eigenvector in each iteration, 
as the poisoned samples may be projected in different 
orthogonal directions \cite{DBLP:conf/sp/ChoudharyKS24}.
Recomputing the dominant eigenvector repeatedly removes these carefully placed outliers. This removal process is repeated until either all $n\epsilon$
furthest points are removed, or $||\Sigma_{X}||_2$ falls below
$\Gamma$. 

Finally, note that the adversary can always place corrupted points close to the clean points, making the clean and corrupted points statistically indistinguishable. As such, any filtering algorithm could remove some clean points, including this one. But intuitively, removal of a small number of such clean points does not affect the mean of the remaining clean points by much. Prior work defines a "$(\epsilon,\delta)$-stability" condition over the clean sample distribution to capture it, which is stated formally later in Section~\ref{subsection:prelim_assumption}. Ours and prior experiments confirm that this assumption holds practically.



\subsection{Practical Drawbacks of Prior Works} 
\label{subsection:practical_issue}
Practical implementations of the original filtering algorithm
are encumbered by computational and practical constraints below:
\begin{enumerate}
    \item \textbf{Precomputing the threshold $\Gamma$}. 
    The original filtering algorithm assumes that $||\Sigma_S||_2$ is known. 
    Estimating $\Sigma_S$ is difficult since it is difficult to obtain unpoisoned clean samples. Furthermore, if threshold $\Gamma$ is conservatively chosen so as to avoid filtering out inliers, it enables an adaptive attack that places points right below $\Gamma$ (e.g. HiDRA~\cite{DBLP:conf/sp/ChoudharyKS24}).
    
    \item \textbf{Eigendecomposition of $\Sigma_X$}. Exact and
    approximate eigendecomposition
    of $\Sigma_X$ involves $O(d^3)$ and $O(d^2 \log d)$ operations, respectively,
    in each of the $O(n\epsilon)$ iterations (See \cite{lay2003linear}, Chapter 5 for more details). Given that most ML models involve aggregating over high-dimensional vectors, this is
    computationally expensive. 
\end{enumerate}

We tackle these challenges to provide a practical solution with provable security guarantees of a strong robust aggregator. 


\section{\textsc{RandEigen}: Our Proposed Robust Aggregator} \label{section:proposed_randeigen}
We propose \randeigen, an efficient randomized strong robust aggregator that performs robust aggregation on \( n \) samples, each of \( d \)-dimensions, with a computational complexity of \( \tilde{O}(nd) \) operations, and
eliminates the dependence on a predetermined threshold derived from the estimated clean sample covariance matrix.
The pseudocode of \randeigen~  is given in Algorithm \ref{alg:randeigen}. 
For convenience, we summarize the notations used in Sections \ref{section:proposed_randeigen} and \ref{section:analysis} in Table \ref{table:randeigen_notation}.

\begin{table}[]
\caption{Notation used in Sections \ref{section:proposed_randeigen} and \ref{section:analysis}. 
JL
refers to the Johnson–Lindenstrauss Transformation. }
\label{table:randeigen_notation}
\scalebox{0.8}{
\begin{tabular}{|c|l|}
\hline
\textbf{Notation} &
  \multicolumn{1}{c|}{\textbf{Description}} \\ \hline
X &
  \begin{tabular}[c]{@{}l@{}}$n \times d$ matrix, consisting of\\ $n$ samples of $d$-dimensions each.\end{tabular} \\ \hline
  $S, T$ &
  Subsets of clean and poisoned samples resp. in X\\ \hline
  $\mu_X, \mu_S$ &
  Empirical mean of $X$ and $S$ resp. \\ \hline
    $\Sigma_X, \Sigma_S$ &
  Empirical Covariance matrix of $X$ and $S$ resp. \\ \hline
$\epsilon_{JL}$ &
  Approximation Error in JL transformation. \\ \hline
$k := \frac{\log d}{\epsilon_{JL}}$ &
  Number of dimensions for JL. \\ \hline
$\epsilon_{P}$ &
  Error rate of power iteration. \\ \hline
Y &
  \begin{tabular}[c]{@{}l@{}}Reduction of $X$ from $d$ to $k$\\ dimensions using JL transformation, \\ represented by a $n \times k$ matrix,\\ where $k = \Omega(\log n/\epsilon_{JL})$.\end{tabular} \\ \hline
$\mathcal{N}_k$ &
  \begin{tabular}[c]{@{}l@{}}$k$-dimensional multivariate \\ normal distribution.\end{tabular} \\ \hline
\begin{tabular}[c]{@{}c@{}}$||x||_p$,\\ $x$ is a vector\end{tabular} &
   $\ell_p$ norm of $x$. \\ \hline
   $\stackrel{f}{=}$ &
  Convergence check (need for floating-point errors). \\ \hline
\end{tabular}
}
\end{table}

\subsection{Overview}

\begin{algorithm}[t]
\caption{Pseudo-code describing RandEigen}
\label{alg:randeigen}
\setlength{\textfloatsep}{0pt}

\begin{algorithmic}[1] 
\REQUIRE Fraction of corruption $\epsilon$, 
set of $\epsilon$-corrupted updates represented by $n \times d$ matrix $X = \{x_1, \dots, x_d\}$, Error of JL approximation $\epsilon_{JL}$, Error of Power Iteration $\epsilon_{P}$
\ENSURE Robust aggregate $\mu = \{\mu_1, \dots, \mu_d\}$, where $\mu_i$ is the robust aggregate of the $i$-th dimension
\STATE $k := \log(d)/\epsilon_{JL}^2$
\STATE Construct $d \times k$ random matrix $A := \{a_1, \dots, a_d\}$, where $a_i \sim \mathcal{N}_k(0, 1/k)$
\STATE $Y := XA$
\FOR{$j=1$ {\bfseries to} $j=  2 n \cdot \epsilon$} 
    \STATEx \quad $\triangleright$ estimated dominant eigenvector/value
    \STATE $\lambda_{curr}, v := POWER-ITERATION(Cov(Y), \epsilon_{p})$ 
    \STATEx \quad $\triangleright$  Eigenvalue has converged
    \IF{$j > 0$ \bf{and} $\lambda_{curr} \stackrel{f}{=}  \lambda_{old}$} \STATE break
    \ELSE \STATE $\lambda_{old}  = \lambda_{curr}$
    \ENDIF
    \STATE $\mu_y = \frac{1}{|Y|} \sum_{y \in Y} y$
    \STATE  $\mathcal{P} = \{p_1, \dots, p_n\}$, where $p_i = |\langle y_i, v \rangle -  \langle \mu_y, v \rangle|, y_i \in Y$
    \IF{$|Y_i| \leq (1-5\epsilon)n$} break
    \ENDIF
    \FOR {$i= 1$ {\bfseries to} $ |Y|$} 
        \STATE Remove $y_i$ from $Y$ with probability $p_i / \max \mathcal{P}$
        \IF{$y_i$ is removed}
            \STATE Remove $x_i$ from $X$
        \ENDIF
    \ENDFOR
\ENDFOR
\State \textbf{return} $\mu$ = Dimension-wise average of $X$
\end{algorithmic}
\end{algorithm}

To the best of our our knowledge, \randeigen~is the first strong 
robust aggregator which achieves information-theoretic
optimal bias in a quasi-linear runtime, which solves the issues
highlighted in Section \ref{subsection:practical_issue}. We achieve this using the following $2$ key ideas:
\begin{enumerate}
    \item \textbf{Convergence of eigenvalues as a stopping criteria.} 
    \randeigen~ eliminates the need for a precomputed threshold $\Gamma$, and instead, replaces the
    threshold based-stopping criteria 
    with a convergence condition based on the eigenvalues of successive iterations. This stems from the following observation: When the iteratively filtered set $X$ starts to have points whose covariance is getting close to the benign variance (as in the original algorithm), 
    {\em the corresponding dominant eigenvalue starts to converge to some value} as well. We formalize this intuition in Theorem \ref{thm:bias_revised}. We can thus detect such convergence and safely stop iterating.
    \item \textbf{Dimensionality Reduction.} The main computational botteneck in existing
    practical strong robust aggregators stems from repeated eigendecomposition over $\Sigma_X$,
    which requires $\Omega(nd^2)$ operations, which impedes the practicality of such aggregators
    in high dimensions. \randeigen~ reduces this to  
    $O((\log^2 d) \log \log d)$ operations, 
    using the following steps: 
    \begin{itemize}
        \item Reduce dimensions from $d$
        to $O(\log d)$ through the Johnson-Lindenstrauss (JL) transformation; 
        \item Compute eigenvector approximations on the $O(\log d)$-dimensional vectors using the power iteration method in $O((\log^2 d) \log \log d)$ operations.
    \end{itemize}
     
\end{enumerate}

The JL transform is a randomized projection of a $d$-dimensional vector down to $O(\log d)$ dimensions such that the length of a vector ($\ell_2$ norm) is preserved after projection on expectation. The power iteration method is a way of approximately converging to the largest eigenvector of a matrix. Our theoretical analysis in Section \ref{section:analysis} confirms that the approximation error is very small.

The final $\randeigen$ algorithm is shown in Algorithm \ref{alg:randeigen}. It retains the main steps of Algorithm~\ref{alg:filter_original}, but incorporates both these insights as modifications to the existing randomized filtering algorithm. 
Specifically, Line 3 performs the dimensionality reduction step.
Line 6 uses power iteration for eigendecomposition. Line 7 and 14
uses the convergence-based stopping criteria explicitly and implicitly respectively. Line 13 and 17-20 filter vectors with reduced dimensionality. The rest of the algorithm remains the same. We explain each of these ideas in detail next.

\subsection{Eigenvalue Convergence for Stopping}

We remove the dependence
on the threshold $\Gamma$, by replacing the stopping condition
of the original filtering algorithm to the following: 
iterative filtering must continue until the dominant eigenvalue converges,
or until the number of iterations reaches the number of poisoned samples,
whichever occurs first. 
To illustrate this, we modify the \texttt{Randomized Filtering}
algorithm to incorporate this stopping
condition, which is reflected in Line 
\ref{algo:stop} of Algorithm \ref{alg:filter_modified}. 
Algorithm \ref{alg:filter_modified} is useful for formal analysis; it is bridge between our final \randeigen~( Algorithm~\ref{alg:randeigen}) and original filtering (Algorithm~\ref{alg:filter_original}).

Our formal proof that Algorithm~\ref{alg:filter_modified}  estimates 
$\mu_S$ up to the information theoretic optimal bias of $O(\sqrt{||\Sigma_S||_2})$, 
is in 
Theorem \ref{thm:bias_revised}.

\setlength{\textfloatsep}{0pt}
\begin{algorithm}[t]
\caption{\texttt{Randomized Filtering} modified to include \\ only the stopping criterion from \randeigen~(used in the analysis).}
\label{alg:filter_modified}
\begin{algorithmic}[1] 
\REQUIRE Fraction of corruption $\epsilon$, 
set of $n$ $\epsilon$-corrupted updates represented by $n \times d$ matrix $X = \{x_1, \dots, x_d\}$
\ENSURE Robust aggregate $\mu = \{\mu_1, \dots, \mu_d\}$, where $\mu_i$ is the robust aggregate of the $i$-th dimension

\FOR{$j=1$ {\bfseries to} $2 n \cdot \epsilon$} 
    \STATEx \quad $\triangleright$  Obtain dominant eigenvalue/vector
    \STATE $\lambda_{curr}, v := EIGENDECOMPOSITION(Cov(X))$ 
    \STATEx \quad $\triangleright$  Eigenvalue has converged
    \IF{$j > 1$ \bf{and} $\lambda_{curr} \stackrel{f}{=}  \lambda_{old}$} \label{algo:stop}
         break
    \ELSE
        \STATE $\lambda_{old}  = \lambda_{curr}$
    \ENDIF
    \STATE $\mu_x = \frac{1}{|X|} \sum_{x \in X} x$ 
    \STATE  $\mathcal{P} = \{p_1, \dots, p_{|X|}\}$, where $p_i = |\langle x_i, v \rangle -  \langle \mu_x, v  \rangle|$ 
    \IF{$|X_i| \leq (1-5\epsilon)n$} break
    \ENDIF
    \FOR {$i= 1$ {\bfseries to} $ |X|$} 
        \STATE Remove $x_i$ from $X$ with probability $p_i / \max \mathcal{P}$
    \ENDFOR
\ENDFOR
\State \textbf{return} $\mu$ = Dimension-wise average of $X$
\end{algorithmic}
\end{algorithm}

\subsection{Randomized Dimensionality Reduction}

To obtain Algorithm~\ref{alg:randeigen} from Algorithm~\ref{alg:filter_modified}, the only other change is that for computing the eigendecomposition approximately.

We represent the $n$ $d$-dimensional samples as a $n \times d$ matrix $\X$, and
perform dimensionality reduction by randomly
projecting $\X$ to $k$ dimensions, for some $k = \log(d)/\epsilon_{JL}^2$, for
some predetermined $\epsilon_{JL} \in (0, 1)$. 
This is the Johnson-Lindenstrauss (JL) transform~\cite{JL_paper} and is constructed as follows:
Let $\mathbf{X}$ denote the original $n \times d$ matrix, and $A$ is a $d \times k$ matrix, 
where each entry in $A$ are independently and identically distributed
(i.i.d.) $\mathcal{N}(0,1)$ random variables. Then, $Y = (1/\sqrt{k}) XA$ is a random projection
of $X$ to $k$ dimensions. 
By the Johnson–Lindenstrauss lemma \cite{JL_paper}, if $k = \Omega(\log n/\epsilon_{JL}^2)$ and for all 
$x_1, x_2 \in \mathbf{X}$, $||x_1 - x_2||_2 /||x_1A -x_2 A||_2 \in (1- \epsilon_{JL}, 1+\epsilon_{JL})$ with high probability. Informally, the JL lemma states that random
projection preserve pairwise distance within a factor of $1 \pm \epsilon_{JL}$, refer to 
\cite{DBLP:journals/corr/abs-2103-00564} for more details. This random projection involves 
$O(ndk) = \tilde{O}(nd)$ operations.
In our experiments, we use $\epsilon_{JL} = 0.1$, as it works well. 

In Theorems \ref{thm:rs_ujl_u} and \ref{thm:dist_jl}, we formally establish that a linear mapping exists between the exact and expected dominant eigenvectors  of $X^T X$ and $Y^T Y$, respectively. Furthermore, we show that the projections of samples from 
$Y$ onto its dominant eigenvector provide a good approximation of the corresponding projections on $X$. 

\subsection{Computing Dominant Eigenvectors} \label{subsection:power_iter}
We apply the power iteration
algorithm \cite{power_iteration} to estimate the dominant eigenvalue and eigenvector of $\Sigma = \text{Cov}(\XJL)$ (which is a $k \times k$
matrix). 
It is known that to estimate the dominant eigenvalue of a $k \times k$ matrix up to an error factor of
$\epsilon_P \in (0, 1)$, at least $\left|\frac{\log(4k)}{2 \log(1-\epsilon_P)}\right|$
iterations of the power iteration method are needed \cite{BAI202129}. Computing $\Sigma$
requires $O(k^3) \in O((\log(d))^3)$ 
 operations.
The power iteration algorithm is performed as follows. 
\begin{enumerate}
    \item Pick a random starting point $v_0 \sim \mathcal{N}_k (0, I)$.
    \item Fix the number of iterations as $N = \left|\frac{\log(4k)}{2 \log(1-\epsilon)}\right|$
    \item For $i = 1, \dots, N$ iterations, compute $v_i = \XJL^T \XJL v_{i-1}$.
\end{enumerate}
Steps (1) - (3) of the power iteration algorithm requires $O(N k^2) = O((\log^2 d) \log \log d)$ operations. In our
evaluations, $\epsilon_{P} = 0.1$ works well. In Theorem \ref{thm:power_iteration_error},
we prove that the above choice of $N$ yields a dominant eigenvector estimate with an error factor of $\epsilon_P$, and that the error of the approximated dominant eigenvector is $O((\lambda_2/\lambda_1)^{2N})$.

\subsection{Stopping Criterion} \label{subsection:filtering_preprocess}

Given its approximated dominant eigenvector $u$, each $y_i \in Y$ in \randeigen~(Algorithm \ref{alg:randeigen}) is assigned a removal probability
$p_i$, as follows,
\begin{enumerate}
    \item Calculate $\tilde{p}_i = |\langle y_i - \mu_y, u \rangle|$, the projection of $y_i - \mu_y$ on $u$.
    \item Set $p_i = 1 - \tilde{p}_i/p_{max}$, where $p_{max} = \max_{j=1, \dots, n} \tilde{p}_j$.
    \item Remove point $x_i$ with probability $p_i$, by genarating $u_i \sim \mathcal{U}$ and removing $p_i$ if $u_i \geq p_i$, 
    where $\mathcal{U}$ is uniform over $[0, 1]$. 
\end{enumerate}

These steps are the same as the original filtering algorithm given in Algorithm~\ref{alg:filter_original}. The main change we make is in the stopping criteria. These steps repeat until one of the conditions below are met:
\begin{enumerate}
    \item The number of iterations has reached $2n\epsilon$,
    \item If the dominant eigenvalue converges between successive iterations (Line 7, Algorithm~\ref{alg:randeigen}), or
    \item When the number of samples remaining drops below $(1-5\epsilon)n$ (Line 14, Algorithm~\ref{alg:randeigen}). 
\end{enumerate}

The first condition is the same as the original filtering algorithm. The algorithm removes at least one point in each iteration with probability $1$ and the $2n\epsilon$ iterations suffice to remove all malicious points that are significantly far away from where the benign points concentrate. The main crux of why the original Algorithm~\ref{alg:filter_original} works is because of the stopping criteria that the dominant eigenvector $\lambda_{curr}$ (Line 3) is below the threshold $\Gamma$, which is in $O(||\Sigma_S||_2)$. Our algorithm introduces two new conditions (2) and (3) for stopping. Condition (3) is to ensure that a lot of benign points do not get filtered out in any iteration. This is implicitly dealt with by $\Gamma$ thresholding in the original Algorithm~\ref{alg:filter_modified} but we must explicitly deal with it since we eliminate explicit thresholding. We later prove that condition (3) implies that the bias has become small enough  and it is safe to stop without filtering out any more points. The correctness of \randeigen~(Algorithm~\ref{alg:randeigen}), therefore, largely rests on the following observation: The dominant eigenvalue starts converges to a value if and only if the bias in the mean of sample vectors remaining after some iteration has become small, i.e, near the optimal O($\sqrt{||\Sigma_S||_2}$). This is the convergence condition (2).

The $\stackrel{f}{=}$ on Line $7$ is a convergence check, which computes the difference between the previous and current eigenvalue and checks if it is small enough. To determine if $\lambda_{i} \stackrel{f}{=} \lambda_{i+1}$, we use $\frac{|\lambda_{i+1} - \lambda_i|}{|\lambda_i|} \leq 10^{-5}$ as the threshold to account for floating-point errors. It is empirically determined.


\section{Theoretical Proofs for \randeigen~} \label{section:analysis}

We formally analyze the correctness of \randeigen. 
We do this in two steps. Algorithm~\ref{alg:filter_modified} incorporates only the stopping criterion of \randeigen~into the original filtering in Algorithm~\ref{alg:filter_original} to show that introducing this change does not impact correctness. We then show our eigendecomposition approximation used in the final \randeigen~algorithm are close to their exact versions on expectation, which bridges the remaining gap in correctness of Algorithm~\ref{alg:filter_modified} and our Algorithm~\ref{alg:randeigen}.


\subsection{Preliminaries and Assumptions} \label{subsection:prelim_assumption}

Prior works on iterative filtering~\cite{DBLP:journals/corr/abs-1911-05911, DBLP:conf/icml/DiakonikolasKK019} assume that the clean sample set $S$ satisfy a regularity condition, known as $(\epsilon, \delta)$-stability.
The definition of $(\epsilon, \delta)$-stability 
is given below in Definition \ref{dfn:epsilon_delta_stable}.
\begin{definition}[\emph{$(\epsilon, \delta)$-stability}] \label{dfn:epsilon_delta_stable}
    A finite set $S \subset \mathbb{R}^d$ 
    with empirical mean $\mu_S$ and empirical covariance $\Sigma_S$ is $(\epsilon, \delta)$-stable if for every unit vector 
    $v \in \mathbb{R}^d$, and every $S' \subseteq S$ with $|S'| \geq (1-\epsilon)|S|$,
    the following conditions hold,
    \begin{enumerate}
        \item $\left| \frac{1}{|S'|} \sum_{s \in S'} v\cdot(s-\mu_S) \right| \leq \delta $, and
        \item $ \left| \frac{1}{|S'|} \sum_{s\in S'} (v\cdot(s-\mu_S))^2 - ||\Sigma_S||_2 \right| \leq \frac{\delta^2}{\epsilon}.$
    \end{enumerate}
\end{definition}

The $(\epsilon, \delta)$-stability condition holds with high probability if the underlying
distribution from which $S$ is sampled has finite mean and covariance, and $|S|$ is large enough. Prior works  explain why the finiteness assumption is minimal---if we discount it, one can construct pathological distributions from which a sample $S$ has large bias resulting from $\epsilon$ fraction of its elements even before corruption~\cite{diakonikolas2023algorithmic}. 
The following formalizes our assumption about $S$.

\begin{assumption}
\label{ass:finite_mean_var}
We assume that the corrupted set of samples $X$ is constructed by corrupting an $\epsilon \leq 1/12 $ proportion 
of samples in $S$, where $S$ is a $(5\epsilon, \delta)$-stable set for
$\delta = \sqrt{20||\Sigma_S||_2}$.
\end{assumption}

Our choice of $\epsilon$ and $\delta$ is nearly the same as prior work~\cite{diakonikolas2019robust,DBLP:conf/sp/ChoudharyKS24}.

\subsection{Analysis without Dimensionality Reduction}
There are $3$ possible stopping criterions for Algorithm \ref{alg:filter_modified}, as described in Section \ref{subsection:filtering_preprocess}. Our Theorem \ref{thm:bias_revised} shows that when any of the stopping criterion is met, the maximum bias of remaining samples (w.r.t the set of clean samples) in any direction is upper bounded by $O(\delta)$, as desired.



\begin{theorem} \label{thm:bias_revised}
    Let $S$ be a $(5\epsilon, \delta)$-stable set, where $\epsilon \leq 1/12$, 
    $\delta = \sqrt{20} \sqrt{||\Sigma_S||_2}$, and $X$ is constructed by corrupting an $\epsilon$ proportion of $S$. 
    Suppose the main loop in Algorithm \ref{alg:filter_modified} terminates after $\tau \leq 2n\epsilon$ iterations with $X_{\tau+1}$  as samples left at Line 15.  Then, on expectation,
        $||\mu_{X_{\tau+1}} - \mu_{S}||_2 \leq \beta \sqrt{||\Sigma_S||_2}$, for $\beta = \sqrt{20} \left(\frac{2}{\epsilon} + 2\right)$. 
\end{theorem}
Due to space limits, we defer the exact proof for Theorem \ref{thm:bias_revised} to Appendix~\ref{appendix:thm}, and present a sketch here. 
\begin{proof}[\emph{Proof Sketch}]
    We partition $X$ into an inlier and outlier set $S'$ and $T'$ respectively, where $S'$ consists of points which are at most $\delta' := 2\frac{\delta}{\epsilon}$ away from $\mu_S$ (projected along any unit vector). The precise definition of inliers and outliers is given in Definition \ref{definition:outlier}. The motivation behind this partitioning is that there could be outliers within $S$ which naturally deviate significantly from $\mu_S$---we bound that there are at most $\epsilon n$ such points (Lemma \ref{lemma:max_deviation_mu_s}, Appendix~\ref{appendix:thm})---and there are at most $n\epsilon$ corrupted samples. So, $|T'| \leq 2 \epsilon n$.

    Let $X_i$ be the samples remaining at the start of the $i$-th iteration, $S'_i = X_i \cap S'$ and $T_i' = X_i \cap T'$. In Lemma \ref{lemma:expected_removal} of Appendix~\ref{appendix:thm}, we prove that if $|v\cdot(\mu_{S_i'} - \mu_{X_i})|  > \delta'$, on expectation in each iteration of the main loop, more outlier samples will be removed compared to inliers. Since each iteration removes at least one sample, and outliers deviate more from the mean than inliers, at least one outlier sample is removed. Furthermore, we show that the proportion of outliers is decreasing on expectation (Lemma \ref{lem:martingale}, Appendix~\ref{appendix:thm}). 

    Let $\lambda_i$ denote the dominant eigenvalue of $\Sigma_{X_i}$. Lemma~\ref{lemma:loop_invariant} in Appendix~\ref{appendix:thm} proves that the following loop invariant holds for all iterations $i \leq \tau$: $\lambda_i$ is a decreasing sequence on expectation, $|X_i| \geq (1-5\epsilon)n$ with probability $1-\exp(-\Omega(n\epsilon))$, and $i \leq 2n\epsilon$. 
    

    Theorem \ref{theorem:max_bias} of  Appendix~\ref{appendix:thm} combines the above Lemmas to show that when Algorithm~\ref{alg:filter_modified} terminates through any one of its stopping conditions, either all outliers have been removed or the mean of remaining samples is within $\delta'$ of that of inliers remaining. Specfically, $||\mu_{X_{\tau+1}} - \mu_{S_{\tau+1}}||_2 \leq \delta'$ on expectation taken over the coin flips in the last iteration $\tau$. 

    Using the above, Theorem \ref{thm:bias_revised} establishes that after the termination of Algorithm \ref{alg:filter_modified} $||\mu_{X_{\tau+1}} - \mu_{S}||_2 \leq \beta \sqrt{||\Sigma_S||_2}$ on expectation. 
\end{proof}

For a fixed constant $\epsilon$, the bias of Algorithm~\ref{alg:filter_modified} is thus $O(1)\sqrt{||\Sigma_S||_2}$.

\subsection{Analysis with Dimensionality Reduction} \label{section:dimensionality_reduction}

We first state some of the relationships between the dominant eigenvectors of $\X$ and 
$\XJL$, before and after the JL transform as follows:
\begin{enumerate}
    \item (Theorem \ref{thm:rs_ujl_u}) There exists a linear mapping between the dominant eigenvector of $\XJL$ and the dominant eigenvector of $\mathbf{X}$.
    \item (Theorem \ref{thm:dist_jl}) If $k = \Omega(\log (d))$ , the difference 
    between the projection of $\X$ and $\XJL$ on
    their respective dominant eigenvectors is small, with probability at least $1 - O\left( \frac{1}{d^{0.225}}\right)$. In Corollary \ref{corr:prob_diff}, we apply Theorem \ref{thm:dist_jl} to prove that the difference in removal probabilities computed in Algorithm \ref{alg:randeigen} and Algorithm \ref{alg:filter_modified} is upper bounded by a small constant, with the same probability. 
\end{enumerate}

We will state Theorems \ref{thm:rs_ujl_u} and \ref{thm:dist_jl} here, and their proofs are deferred to
Appendix \ref{appendix:proof}. 

\begin{theorem} \label{thm:rs_ujl_u}
Given a $n \times d$  matrix $\X$, and its corresponding $k$-dimensional Johnson-Lindenstrauss transform, 
$\XJL = (1/\sqrt{k}) \X A$, where $A$ is a $d \times k$ matrix of i.i.d. $\mathcal{N}(0,1)$ random variables. Let $u$
and $v$ denote the dominant eigenvectors of $\X^T \X$ and $\XJL^T \XJL$, respectively. Then, 
it follows that $v = (1/\sqrt{k})  u A$ is the expected eigenvector of $\XJL^T \XJL$, $\mathbb{E} [(1/\sqrt{k}) v A^T] = u$ and the expected dominant eigenvalue of $\XJL^T \XJL$ is $k \lambda_1$, where $ \lambda_1$ is the dominant eigenvalue of $\X^T \X$.
\end{theorem}
\begin{remark}
   To clarify notation, in Algorithms \ref{alg:filter_original} and \ref{alg:filter_modified}, the dominant eigenvector is computed over the empirical covariance matrix $\X^T \X$, where (boldface) $\X$ has each row vector as the deviation of original $x \in X$ from the mean $\mu_X$. Specifically,  $\X = (X - \bf{1}^n\mu_{X})$, where $\bf{1}^n \in \mathbb{R}^{n\times 1}$ is the  vector of all ones and $\mu_X \in \mathbb{R}^{ 1\times d}$. In Algorithm \ref{alg:randeigen}, the same dominant eigenvector is computed over $\XJL^T \XJL$, where $\XJL = (X - \bf{1}^n\mu_{X})A$. 
\end{remark}

\begin{theorem} \label{thm:dist_jl}
    Continuing from the notation in Theorem \ref{thm:rs_ujl_u}, for any $x \in X$, let $f(x) = \frac{1}{k}xA$ and $\hat{x} = (x_1u_1, \dots, x_d u_d)$ denote the components of $x$ after projecting on the dominant eigenvector $u$. Let $P_X = \max \mathcal{P} = \max_{x \in X} |u \cdot (x - \mu_X)|$ (as defined in Algorithm \ref{alg:filter_modified}). 
    Then, $v=f(u)$ is the expected dominant eigenvector of $f(X)$, and
    \[
    \IP\left[ \left| \frac{\hat{x} - \hat{\mu}_X}{P_X} \cdot u - f\left(\frac{\hat{x}- \hat{\mu}_X}{P_X}\right) f(u) \right| \geq \epsilon_{JL}\right] \leq 4\exp\left(- \frac{(\epsilon_{JL}^2 - \epsilon_{JL}^3)k}{4} \right).
    \]
    Using $\epsilon_{JL} = 0.1$ and $k = \frac{\log d}{\epsilon_{JL}^2}$ (which are the values used in Algorithm \ref{alg:randeigen},
    \[
    \IP\left[ \left| \frac{\hat{x}- \hat{\mu}_X}{P_X} \cdot u - f\left(\frac{\hat{x}- \hat{\mu}_X}{P_X}\right) f(u) \right| \geq \epsilon_{JL}\right] \leq \frac{4}{d^{0.225}}.
    \]
\end{theorem}
Theorem \ref{thm:dist_jl} shows that, with high probability, the random projection of any vector $x\in X$ from $d$-dimensional space down to $O(\log d)$-dimensions preserves its component along the expected dominant eigenvector of the lower-dimensional subspace. This fact is used in Corollary \ref{corr:prob_diff} to show that the removal probabilities computed in Algorithm \ref{alg:randeigen} closely approximates those computed in Algorithm \ref{alg:filter_modified}. 
\begin{corollary} \label{corr:prob_diff}
    Continuing from Theorem \ref{thm:dist_jl}, for any $x \in X$ and $y = f(x)$,
    let $p_x$ denote the removal probability computed in Line 12 of Algorithm \ref{alg:filter_modified}, and $p_y$ denote the removal probability computed  in Line 16 of Algorithm \ref{alg:randeigen} respectively when $y$ is projected on the expected dominant eigenvector of $\Sigma_Y$.
    If $k$ and $\gamma = \epsilon_{JL}$ are as defined in Algorithm \ref{alg:randeigen}, the following concentration bound holds:
    \[
    \begin{split}
        \IP\left[|p_x - p_y| \geq \frac{\epsilon_{JL}}{1-\epsilon_{JL}} \right]\leq  \frac{8}{d^{0.225}}.
    \end{split}
    \]
\end{corollary}

\subsection{Dominant Eigenvector Estimation}

In Theorem \ref{thm:power_iteration_error}, 
we show that the the number of iterations needed to estimate the dominant eigenvalue
within $\epsilon_p$ precision, for some $\epsilon_p \in (0,1)$, is $N \geq  \max\left(\left| \frac{\log (4k)}{2 \log (1-\epsilon_p)} \right|, \left|\frac{\log \epsilon_p}{2 \log (\lambda_2 / \lambda_1)} \right| \right)$,\footnote{In our evaluations, we found that taking $N \geq  \left| \frac{\log (4k)}{2 \log (1-\epsilon_p)} \right|$ number of iterations works well.}
and the the $\ell_2$ distance between the estimated and actual dominant eigenvectors converges to $0$ 
at a rate exponential to $2N$.
\begin{theorem} \label{thm:power_iteration_error}
    Given a $k \times k$  matrix $M$, let $\lambda_1, \lambda_2, \dots, \lambda_k$ be the eigenvalues
    associated with eigenvectors $u_1, u_2, \dots, u_k$, where 
    $|\lambda_1| \geq |\lambda_2| \geq \dots \geq |\lambda_k|$, and let $v_N$
    and $\gamma_N$ be the estimate of 
    $u_1$ and $\lambda_1$, respectively, using the power iteration algorithm after $N$ iterations. 
    For any $\epsilon_p \in(0,1)$,
    a necessary condition for $\left| \frac{\gamma_N - \lambda_1}{\lambda_1} \right| < \epsilon_p$
    is $N \geq  \max \left(\left| \frac{\log (4k)}{\log (1-\epsilon_p)} \right|, \left|  \frac{\log \epsilon_p}{2 \log (\lambda_2 / \lambda_1)}\right|\right)$. Furthermore, without loss of generality, 
    assume $u_1$ and $v_N$ are normalized, such that $||u_1||_2 = ||v_N||_2 = 1$. Then, 
    $||v_N- u_1||_2 = O\left(\left|\frac{\lambda_2}{\lambda_1}\right|^{2N}\right)$
\end{theorem}
The proof of Theorem \ref{thm:power_iteration_error} is deferred to Appendix \ref{appendix:proof}.

\subsection{Correctness of \randeigen}

We have previously shown via Theorem~\ref{thm:bias_revised} that Algorithm~\ref{alg:filter_modified} is a robust aggregator with near-optimal bias. We now show that our final \randeigen~(Algorithm \ref{alg:randeigen}) also satisfies the same.

\begin{theorem}
    If Algorithm \ref{alg:filter_modified} is a robust aggregator admiting bias at most $\delta$ with 
    high probability, 
    then \randeigen~(Algorithm \ref{alg:randeigen}) is a robust aggregator 
    the same bias on expectation. 
\end{theorem}
\begin{proof}
    The key differences between Algorithm \ref{alg:randeigen} and Algorithm \ref{alg:filter_modified} lies in Lines 1-3 of Algorithm \ref{alg:filter_modified}, where we performed a JL transformation on
    $X$, and Line 6, where we used power iteration to yield an efficient approximation
    of the dominant eigenvector. 

    Consider an original sample $x \in \X$
    and its corresponding JL transformed sample $y \in \XJL$, and 
    let $u$ and $v$ denote the dominant eigenvectors of $\X$ and the expected dominant vector of $\XJL$, respectively.
    In Corollary \ref{corr:prob_diff}, we showed that each of the removal probabilities computed in computed in Line 16 of Algorithm \ref{alg:randeigen}
    approximates those computed in Line 12 of Algorithm \ref{alg:filter_modified} 
    closely, with probability at least $1-\frac{8}{d^{0.225}}$. 
    Next, we showed in Theorem \ref{thm:power_iteration_error} that the minimum number
    of iterations needed to ensure that the dominant eigenvalue is estimated up to
    a factor of $\epsilon_P$ is at least $N = O(\log \log d)$. 

    Therefore, using the above facts, Algorithm \ref{alg:randeigen} uses eigenvalues which closely approximates the exact quantities needed to compute removal probabilities in Line 12 of Algorithm \ref{alg:filter_modified} and the stopping conditions (as given in Theorem \ref{thm:dist_jl}) on expectation. 
\end{proof}




\section{Experiments} \label{section:expt}

We aim to answer the following questions experimentally:
\begin{enumerate}
    \item Does \randeigen~run in the theoretically expected quasi-linear time on real-world machine learning (ML) tasks?
    \item Does \randeigen~affect the performance of the trained ML models in the absence of any poisoned samples?
    \item How effective is \randeigen~against the three types of poisoning attacks?
    \item How does \randeigen~compare to prior practical implementations of strong robust aggregators? 
\end{enumerate}
\subsection{Experimental Setup} \label{subsection:expt_setup}

We focus on the federated learning setup for evaluation wherein
the participating clients send locally trained model updates to a centralized server for aggregation. This is done for every training step and for a predetermined number of training steps. We consider an $\epsilon$-fraction of the clients to be malicious. These malicious clients, depending on their attack objectives, launch poisoning attacks by sending corrupted model updates during each step. To defend against such attacks, the server employs a robust aggregation method that minimizes the resulting bias during aggregation in each training step. The federated learning setup represents a strong adversary for poisoning attacks since the adversary has direct control over the poisoned samples in every training step.


\paragraph{Attacks.} We evaluate \randeigen~ against $7$ attacks spanning all $3$ types of poisoning objectives in the federated setup. These are:

\begin{itemize}
    \item {\em Backdoor}: Distributed Backdoor Attacks (DBA) \cite{xie2019dba}, Model Replacement Attack (MRA) \cite{bagdasaryan2020backdoor}, and Embedded Poisoning (EP)~\cite{yang-etal-2021-careful};
    \item {\em Untargeted}: Trimmed Mean Attack (TMA) \cite{fang2020local} and  
    High Dimensional attack on Robust Aggregators (HIDRA) \cite{DBLP:conf/sp/ChoudharyKS24}
    \item {\em Targeted}: Model Poisoning Attack (MPA) \cite{DBLP:conf/icml/BhagojiCMC19}
\end{itemize}

The above attacks are representative of state-of-the-art effectiveness against practical implementations of strong robust aggregators, FILTERING and NO-REGRET, which we compare against. We implement the attacks as done in the prior work~\cite{zhu2023byzantine,DBLP:conf/sp/ChoudharyKS24}.
DBA, TMA, MRA, EP, and MPA are all {\em non-adaptive} attacks, which are reasonably well mitigated by the only known practical implementations of strong robust aggregators~\cite{zhu2023byzantine}\footnote{MPA is partially mitigated by FILTERING see~\cite{zhu2023byzantine} for details}. HiDRA is state-of-the-art {\em adaptive} attack strategy that is designed to beat all strong robust aggregators. It is provably near-optimal and has been shown to be highly effective against practical implementations. Hence, we choose to include it in our evaluation.

We begin by evaluating the aforementioned attacks on image classification tasks within the federated learning setup. To provide a more comprehensive analysis, we further extend our evaluation to training language models. Specifically, we consider attacking the training of language models using HiDRA and the Embedded poisoning attack (EP)~\cite{yang-etal-2021-careful}. EP is state-of-the-art backdoor attack for language models and has been shown to effectively reduce the performance of these models in sentiment analysis tasks. Therefore, we evaluate how much \randeigen~can prevent performance loss in sentiment analysis tasks due to these attacks.

Lastly, we also evaluate \randeigen~ on a few attacks for the centralized setup as well as attacks originally designed for editing model parameters post-training, adapted to the in-training setup. These evaluation results are given in Section~\ref{subsection:eval_centralized} and~\ref{subsection:dfba} respectively.

\paragraph{Models.} We used the CNN models for image classification used in prior evaluations of these attacks~\cite{zhu2023byzantine,DBLP:conf/sp/ChoudharyKS24}
and for sentiment analysis we use Bert~\cite{devlin2018bert} with fine-tuning. The CNN models have roughly $2^{18}$ parameters (dimensions) and Bert has $2^{27}$ parameters. 

We summarize the datasets, models, application tasks, and their corresponding attacks in Table~\ref{table:evaluation_setup}.

\paragraph{Federated learning setup.} For the image classification tasks, we consider a total of $100$ clients in our federated learning (FL) setup, with $100\epsilon$ malicious clients,
where $\epsilon \in \{0, 0.01, 0.05, 0.10, 0.20\}$. We perform the training for $100$ steps (epochs), and the metrics reported are the average of five trials. We observe that it is sufficient for the models to converge with good accuracy without poisoning. For the sentiment analysis tasks, we finetune BERT to achieve good performance. We perform finetuning for a reduced number of clients ($20$) and for $20$ steps, due to limited computational resources available to us. We choose $\epsilon \in \{0, 0.05, 0.10, 0.20\}$ for this task. We observe that this much finetuning is sufficient for convergence and to achieve close to the accuracy originally reported.

\begin{table}[!htbp]
\caption{Attacks, datasets, and models used for evaluations} \label{table:evaluation_setup}
\begin{tabular}{|l|l|l|l|l|}
\hline
Task                                                                            & Dataset                                                                        & Model                 & Attack & \begin{tabular}[c]{@{}l@{}}Attack\\ Type\end{tabular} \\ \hline
\multirow{5}{*}{\begin{tabular}[c]{@{}l@{}}Image\\ Classification\end{tabular}} & \multirow{5}{*}{\begin{tabular}[c]{@{}l@{}}MNIST \\ \&\\ F-MNIST\end{tabular}} & \multirow{5}{*}{CNN}  & DBA    & Backdoor                                              \\
                                                                                &                                                                                &                       & MRA    & Backdoor                                              \\
                                                                                &                                                                                &                       & TMA    & Untargeted                                            \\
                                                                                &                                                                                &                       & HIDRA  & Untargeted                                              \\
                                                                                &                                                                                &                       & MPA     & Targeted                                              \\ \hline
\multirow{2}{*}{\begin{tabular}[c]{@{}l@{}}Sentiment\\ Analysis\end{tabular}}   & \multirow{2}{*}{SST2}                                                          & \multirow{2}{*}{BERT} & EP     & Backdoor                                              \\
                                                                                &                                                                                &                       & HIDRA  & Untargeted                                            \\ \hline
\end{tabular}
\end{table}

\subsubsection*{Strong Robust Aggregators.}
Recall that prior theoretical designs of strong robust aggregators require running time $\Omega(nd^2)$ and therefore, do not admit any practically realizable strategy.
We compare \randeigen~against the practical implementations of 
 NO-REGRET and FILTERING (Algorithm 2 and 3 of \cite{zhu2023byzantine}, respectively),
which to the best of our knowledge, are the only two strong robust aggregators for which any practical realizations have been suggested. 
%
To overcome memory constraints, these practical implementations divide the samples 
into disjoint chunks of $C$ dimensions each, which are small enough to be computed on individually
and admit a computational complexity of $O(n d C^2)$.
Note that the quadratic dependence on $C$ limits how large the chunk size $C$ can be made. In our evaluation, we used the default setting of $C = 1,000$, as the time taken (in seconds)
for each training step in SGD training is approximately $35, 77, 112, 172$ and $263$, for $C = 1,000, 2,000, 3,000, 4,000, 5,000$, respectively. 

Note that the chunking trick causes the practical realization of prior strong aggregators to lose their optimal bias properties, as demonstrated by the HiDRA attack~\cite{DBLP:conf/sp/ChoudharyKS24}.

%

\subsubsection*{System Configuration}
We use a linux machine with---Processor: Intel(R) Xeon(R) CPU E5-2620 v4 @ 2.10GHz;
RAM: 256GB;
OS:  Ubuntu 20.04.4 LTS;
GPU: NVIDIA Tesla V100. 

\subsubsection*{Evaluation Metrics}
We employ
the following standard metrics:
\begin{enumerate}
    \item Accuracy (ACC): Test accuracy of the final model;
    \item Time: Average time taken across all aggregation steps;
    \item \textit{[For targeted and backdoor attacks]} Attack success rate (ASR): 
    Proportion of poisoned samples that output the targeted result when tested after training. 
\end{enumerate}

\subsection{\randeigen~Runtime Speedup}

\begin{figure}
    \centering
    \subfloat[Actual runtime for different $d$ \label{subfloat:a}]{\includegraphics[width=0.49\linewidth]{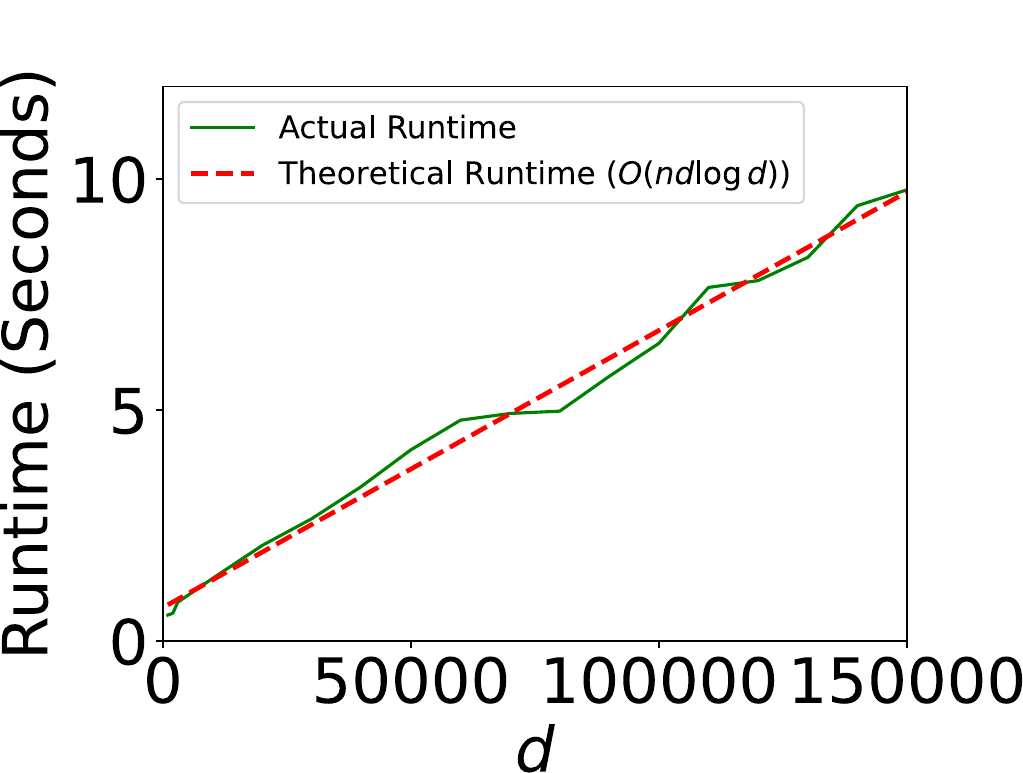}} 
    \subfloat[Ratio between FILTERING and \randeigen~runtime \label{subfloat:b}]{\includegraphics[width=0.49\linewidth]{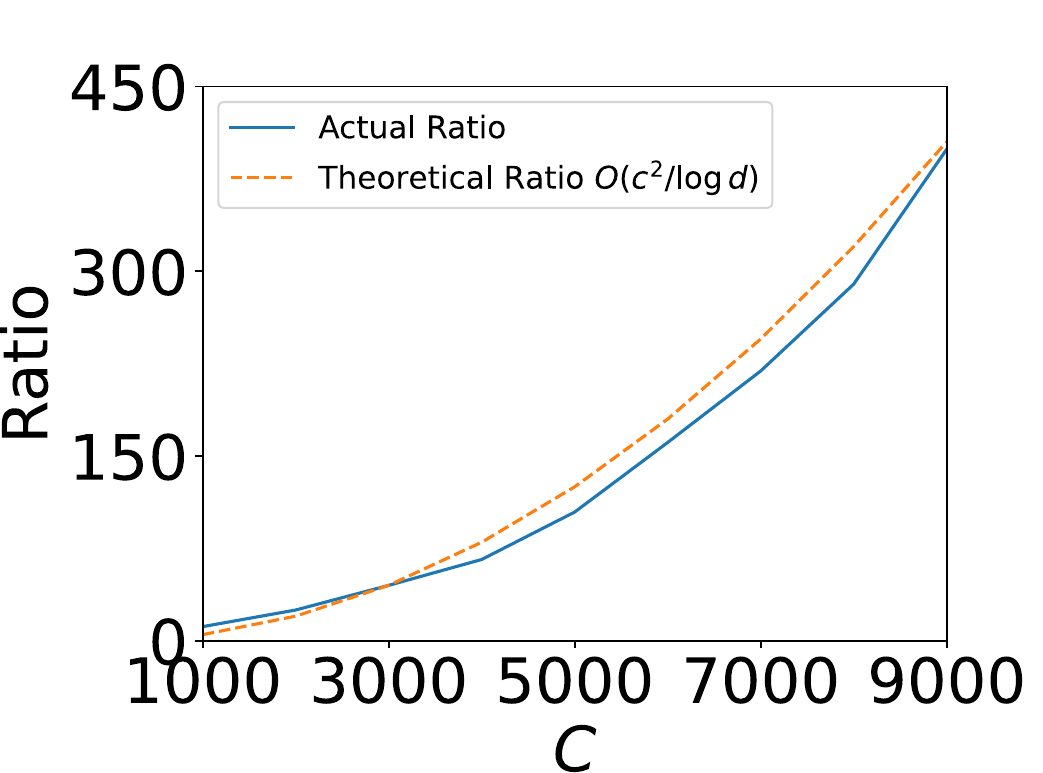}} 
    \caption{Subfigure (\ref{subfloat:a}) illustrates that \randeigen~admits a quasi-linear
    runtime with respect to $d$. Subfigure (\ref{subfloat:a}) illustrates the ratio between the time taken for FILTERING and \randeigen~for various chunk sizes. For both plots, 
    we include the theoretically expected behavior for comparison. 
    }
    \label{fig:ratio_time_filter_randeigen}
    \vspace{-5mm}
\end{figure}

From our theoretical analysis, we expect \randeigen~to have $O(nd\cdot log (d))\approx\tilde{O}(nd)$ running time, if the gradient distributions satisfy our assumptions.
We have experimentally confirmed that the assumptions are mild enough to be satisfied in practice for all of test subject runs. Figure~\ref{subfloat:a} shows the theoretically expected and actual running time plot of \randeigen~with increasing dimensions for a randomly chosen step during training for the CNN model.

The expected computational complexity ratio between FILTERING (Our baselines FILTERING and NO-REGRET have comparable running times) and \randeigen~is
$O\left( n \lceil d/C\rceil C^3 \right) / O(n d \log d) = O(C^2 / \log d)$, where
$C$ is the maximum number of dimensions per chunk. To verify that this ratio grows 
 quadratically in $C$, we computed this ratio for the image classification task, for
different values of $C$. The results summarized in Figure \ref{subfloat:b}, confirms that the empirically observed ratios align with the theoretical ratio.
\randeigen~is about $15\times$ faster even for small dimensional setups of $d=1k$ dimensions, and as theoretically expected, the \randeigen~speedup improves drastically as dimensions increase (e.g. over $300\times$ at about $d=9000$). 

\begin{mdframed}[backgroundcolor=mygray]
\randeigen~runs in quasi-linear time and offers $O(d^2/log(d))$ speedup of prior concrete realizations of strong aggregators.
\end{mdframed}

\subsubsection*{Memory Overhead} 
As $d$ increases the GPU memory saturates for the baseline FILTERING implementation and hence we cannot show it. However, for the practical implementations, at any point in time, FILTERING and NO-REGRET require at most 0.03GB of GPU memory overhead (since they use chunks). It largely corresponds to the amount of memory needed to construct the $C \times C$  covariance matrix, where $C=1,000$. \randeigen~uses dimensionality reduction using JL transform, eliminating the need for chunking. This reduces dimensions from $d$ to $O(\log d)$. We observe that \randeigen~takes up at most 0.05GB of GPU memory, which largely corresponds to space needed for the covariance matrix after JL transformation. Thus, \randeigen~can scale up to much high dimensions on the same GPU (ours has $16$GB) compared to prior works, addressing a key bottleneck for a practical defense.


\begin{table*}[!htbp]
\caption{Model performance and average aggregation time (per training step, in seconds) 
for the image classification task using RandEigen, FILTERING, NO-REGRET, and the arithmetic mean 
aggregators under different types of Byzantine attacks, with $\epsilon = 0.2$. 
The baseline scenario, where there are no malicious clients (i.e., $\epsilon = 0$), is also included for comparison. \randeigen~performs as well as prior defenses in reducing ASR or maintaining ACC (accuracy), with HIDRA showing the largest improvement. }
\label{table:baseline_img}
\resizebox{2\columnwidth}{!}{%
\begin{tabular}{|l|l|l|ccccccccc|}
\hline
\multicolumn{1}{|c|}{\multirow{3}{*}{Type}} & \multicolumn{1}{c|}{\multirow{3}{*}{Attack}} & \multicolumn{1}{c|}{\multirow{3}{*}{Dataset}} & \multicolumn{9}{c|}{Aggregator}                                                                                                                                  \\ \cline{4-12} 
\multicolumn{1}{|c|}{}                      & \multicolumn{1}{c|}{}                        & \multicolumn{1}{c|}{}                         & \multicolumn{3}{c|}{\textbf{\randeigen}}                               & \multicolumn{3}{c|}{FILTERING}                        & \multicolumn{3}{c|}{No Defense} \\ \cline{4-12} 
\multicolumn{1}{|c|}{}                      & \multicolumn{1}{c|}{}                        & \multicolumn{1}{c|}{}                         & ASR             & ACC             & \multicolumn{1}{c|}{Time}          & ASR    & ACC             & \multicolumn{1}{c|}{Time}  & ASR       & ACC       & Time    \\ \hline
\multicolumn{1}{|c|}{\multirow{2}{*}{-}}    & \multirow{2}{*}{No Attack}                   & MNIST                                         & \textbf{-}      & \textbf{94.5}\% & \multicolumn{1}{c|}{\textbf{2.48}} & -      & 93.0\%          & \multicolumn{1}{c|}{37.68} & -         & 94.5\%    & <0.01   \\
\multicolumn{1}{|c|}{}                      &                                              & F-MNIST                                       & -               & \textbf{81.1}\% & \multicolumn{1}{c|}{\textbf{2.33}} & -      & 81.1\%          & \multicolumn{1}{c|}{37.33} & -         & 81.1\%    & <0.01   \\ \hline
\multirow{4}{*}{Backdoor}                   & \multirow{2}{*}{MRA}                         & MNIST                                         & \textbf{5.4}\%  & \textbf{92.4}\% & \multicolumn{1}{c|}{\textbf{2.42}} & 4.9\%  & 92.1\%          & \multicolumn{1}{c|}{36.25} & 19.5\%    & 92.1\%    & <0.01   \\
                                            &                                              & F-MNIST                                       & \textbf{10.0}\% & \textbf{76.1}\% & \multicolumn{1}{c|}{\textbf{2.16}} & 10.4\% & 75.6\%          & \multicolumn{1}{c|}{33.01} & 31.1\%    & 72.8\%    & <0.01   \\ \cline{2-12} 
                                            & \multirow{2}{*}{DBA}                         & MNIST                                         & \textbf{3.1}\%  & \textbf{93.1}\% & \multicolumn{1}{c|}{\textbf{2.40}} & 2.8\%  & 95.0\%          & \multicolumn{1}{c|}{36.25} & 29.7\%    & 89.1\%    & <0.01   \\
                                            &                                              & F-MNIST                                       & \textbf{14.0}\% & \textbf{78.4}\% & \multicolumn{1}{c|}{\textbf{2.78}} & 14.1\% & 79.9\%          & \multicolumn{1}{c|}{39.11} & 34.7\%    & 80.0\%    & <0.01   \\ \hline
\multirow{2}{*}{Targeted}                   & \multirow{2}{*}{MPA}                         & MNIST                                         & \textbf{10.0}\% & \textbf{95.0}\% & \multicolumn{1}{c|}{\textbf{2.32}} & 30.0\% & 93.8\%          & \multicolumn{1}{c|}{32.17} & 100\%     & 95.1\%    & <0.01   \\
                                            &                                              & F-MNIST                                       & \textbf{10.0}\% & \textbf{79.8}\% & \multicolumn{1}{c|}{\textbf{2.38}} & 30.0\% & 79.4\%          & \multicolumn{1}{c|}{39.44} & 100\%     & 73.1\%    & <0.01   \\ \hline
\multirow{4}{*}{Untargeted}                 & \multirow{2}{*}{HIDRA}                       & MNIST                                         & -               & \textbf{93.4}\% & \multicolumn{1}{c|}{\textbf{2.10}} & -      & 9.8\%           & \multicolumn{1}{c|}{37.47} & -         & 10.0\%    & <0.01   \\
                                            &                                              & F-MNIST                                       & -               & \textbf{76.8}\% & \multicolumn{1}{c|}{\textbf{2.21}} & -      & 10.1\%          & \multicolumn{1}{c|}{39.98} & -         & 8.8\%     & <0.01   \\ \cline{2-12} 
                                            & \multirow{2}{*}{TMA}                         & MNIST                                         & -               & \textbf{94.1}\% & \multicolumn{1}{c|}{\textbf{2.57}} & -      & \textbf{94.1}\% & \multicolumn{1}{c|}{31.72} & -         & 14.7\%    & <0.01   \\
                                            &                                              & F-MNIST                                       & -               & \textbf{77.8}\% & \multicolumn{1}{c|}{\textbf{2.41}} & -      & 78.0\%          & \multicolumn{1}{c|}{41.59} & -         & 12.3\%    & <0.01   \\ \hline
\end{tabular}%
}
\end{table*}

\begin{table*}[]
\caption{Model performance and average aggregation time (per training step, in seconds) 
for the sentiment analysis task using \randeigen , FILTERING, NO-REGRET, and the arithmetic mean 
aggregators under different types of Byzantine attacks, with $\epsilon = 0.2$. 
The baseline scenario, where there are no malicious clients (i.e., $\epsilon = 0$), is also included for comparison. \randeigen~performs as well as prior defenses in reducing ASR or maintaining ACC (accuracy), with HIDRA showing the largest improvement.
}
\label{table:baseline_sst}
\begin{tabular}{|c|l|cccccccccccc|}
\hline
\multirow{3}{*}{Type}            & \multicolumn{1}{c|}{\multirow{3}{*}{Attack}} & \multicolumn{12}{c|}{Aggregator}                                                                                                                                                                                                                                                                                                                 \\ \cline{3-14} 
                                 & \multicolumn{1}{c|}{}                        & \multicolumn{3}{c|}{\textbf{\randeigen} }                                                         & \multicolumn{3}{c|}{FILTERING}                                                          & \multicolumn{3}{c|}{NO-REGRET}                                                          & \multicolumn{3}{c|}{No Defense}                                        \\ \cline{3-14} 
                                 & \multicolumn{1}{c|}{}                        & \multicolumn{1}{c|}{ASR}    & \multicolumn{1}{c|}{ACC}     & \multicolumn{1}{c|}{Time} & \multicolumn{1}{c|}{ASR}    & \multicolumn{1}{c|}{ACC}     & \multicolumn{1}{c|}{Time}  & \multicolumn{1}{c|}{ASR}    & \multicolumn{1}{c|}{ACC}     & \multicolumn{1}{c|}{Time}  & \multicolumn{1}{c|}{ASR}      & \multicolumn{1}{c|}{ACC}     & Time \\ \hline
-                                & No Attack                                    & \multicolumn{1}{c|}{-}      & \multicolumn{1}{c|}{\textbf{92.47}\%} & \multicolumn{1}{c|}{\textbf{1.62}} & \multicolumn{1}{c|}{-}      & \multicolumn{1}{c|}{92.86\%} & \multicolumn{1}{c|}{37.68} & \multicolumn{1}{c|}{-}      & \multicolumn{1}{c|}{93.71\%} & \multicolumn{1}{c|}{37.21} & \multicolumn{1}{c|}{-}        & \multicolumn{1}{c|}{94.45\%} & 0.01 \\ \hline
\multicolumn{1}{|l|}{Backdoor}   & EP                                           & \multicolumn{1}{c|}{\textbf{6.58}\%} & \multicolumn{1}{c|}{\textbf{92.40}\%} & \multicolumn{1}{c|}{\textbf{1.75}} & \multicolumn{1}{c|}{9.12\%} & \multicolumn{1}{c|}{91.99\%} & \multicolumn{1}{c|}{35.25} & \multicolumn{1}{c|}{6.60\%} & \multicolumn{1}{c|}{92.43\%} & \multicolumn{1}{c|}{36.47} & \multicolumn{1}{c|}{100.00\%} & \multicolumn{1}{c|}{94.00\%} & 0.01 \\ \hline
\multicolumn{1}{|l|}{Untargeted} & HIDRA                                        & \multicolumn{1}{c|}{-}      & \multicolumn{1}{c|}{\textbf{92.66}\%} & \multicolumn{1}{c|}{\textbf{1.69}} & \multicolumn{1}{c|}{-}      & \multicolumn{1}{c|}{43.66\%} & \multicolumn{1}{c|}{37.47} & \multicolumn{1}{c|}{-}      & \multicolumn{1}{c|}{47.12\%} & \multicolumn{1}{c|}{36.43} & \multicolumn{1}{c|}{-}        & \multicolumn{1}{c|}{44.77\%} & 0.01 \\ \hline
\end{tabular}
\end{table*}

\subsection{\randeigen~in Non-Adversarial Setting}

Our evaluations on both tasks demonstrate that \randeigen~maintains model accuracy in the absence of poisoned samples. 
The results are summarized in the first rows of Tables \ref{table:baseline_img} and \ref{table:baseline_sst}.
Compared to using simple arithmetic mean aggregation, which offers no robustness, \randeigen~loses below $2\%$ in accuracy, which is comparable to FILTERING and NO-REGRET defenses. This is because some clean gradients are outliers and are filtered out, but as we see, their removal has a modest impact on the model accuracy. 

\begin{mdframed}[backgroundcolor=mygray]
    The impact of \randeigen~on accuracy in the absence of poisoned samples is $<2\%$ across all evaluated configurations. 
\end{mdframed}

\subsection{\randeigen~against Poisoning Attacks} \label{subsection:randeigen_fl}

\begin{figure}
    \centering
    \includegraphics[width=0.49\linewidth]{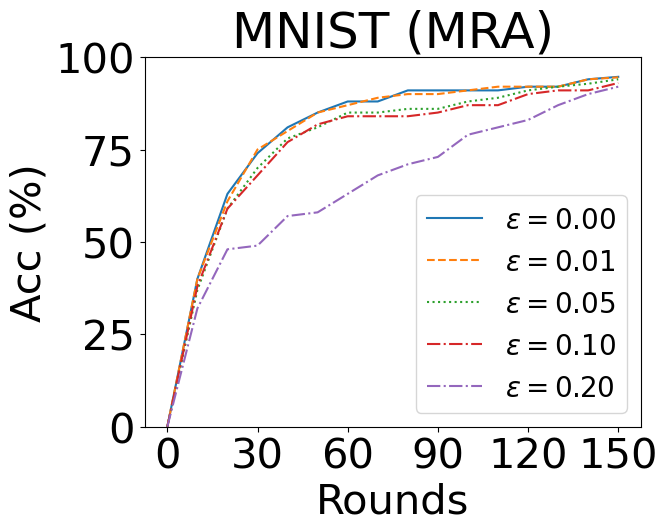}
    \includegraphics[width=0.49\linewidth]{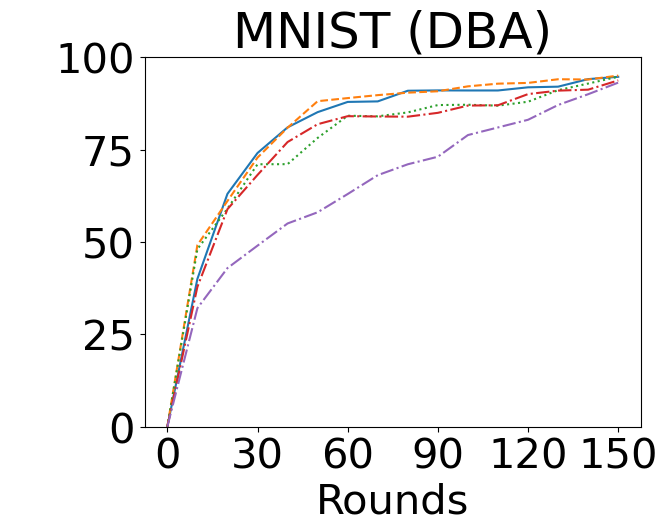}
    \includegraphics[width=0.49\linewidth]{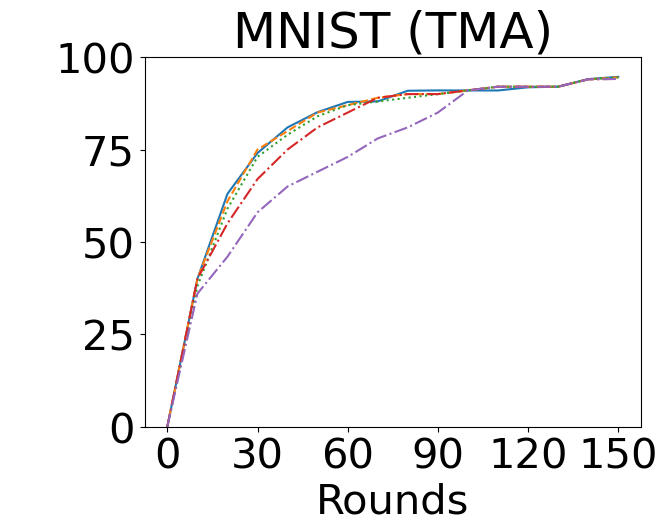}
    \includegraphics[width=0.49\linewidth]{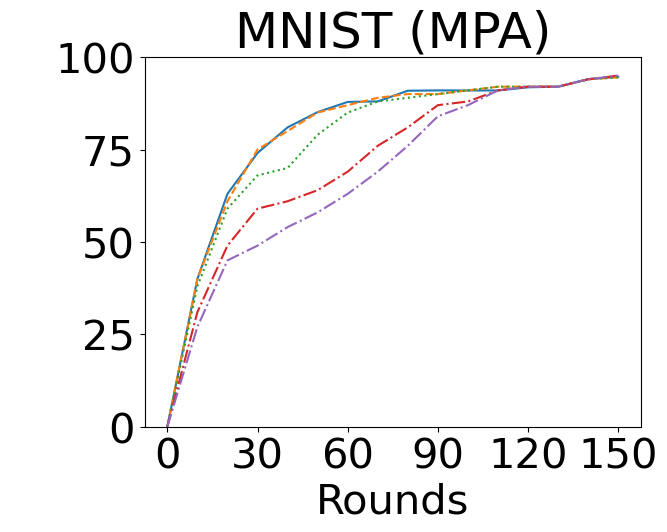}
    \includegraphics[width=0.49\linewidth]{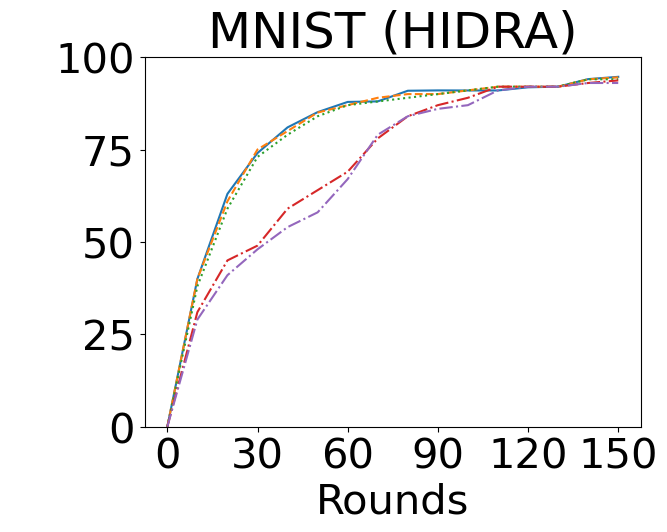} 
    \includegraphics[width=0.49\linewidth]{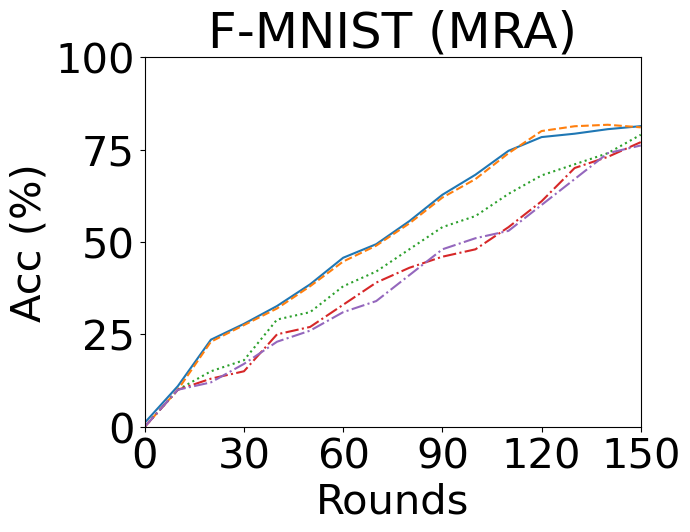}
    \includegraphics[width=0.49\linewidth]{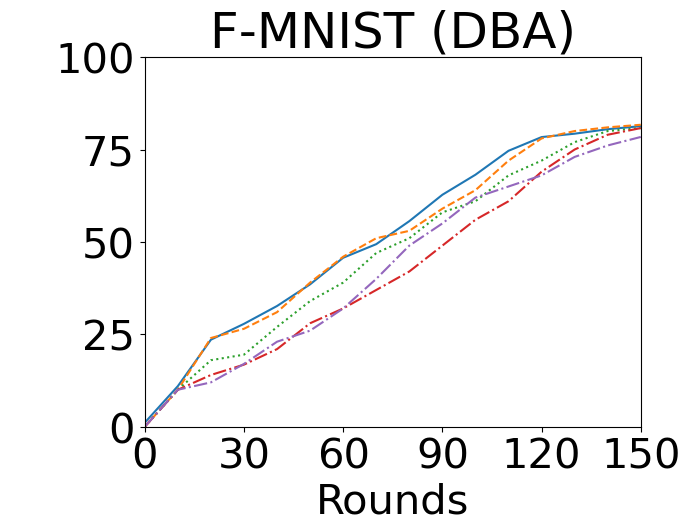}
    \includegraphics[width=0.49\linewidth]{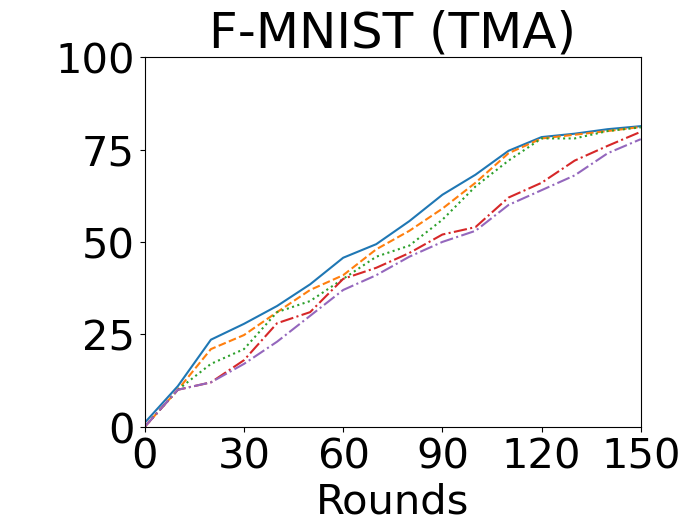}
    \includegraphics[width=0.49\linewidth]{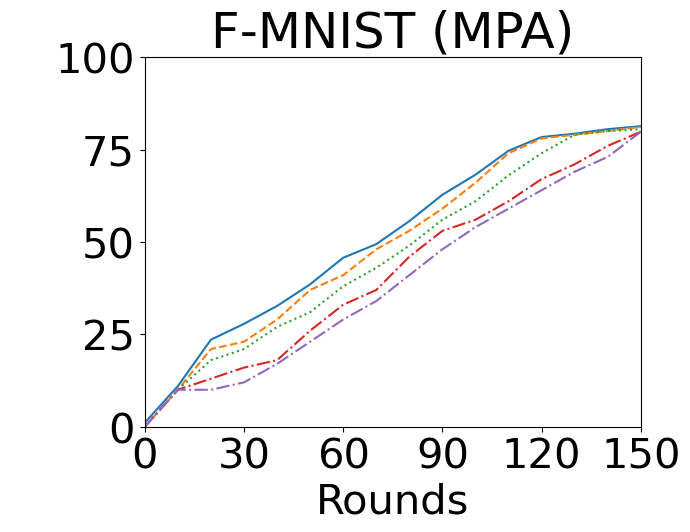}
    \includegraphics[width=0.49\linewidth]{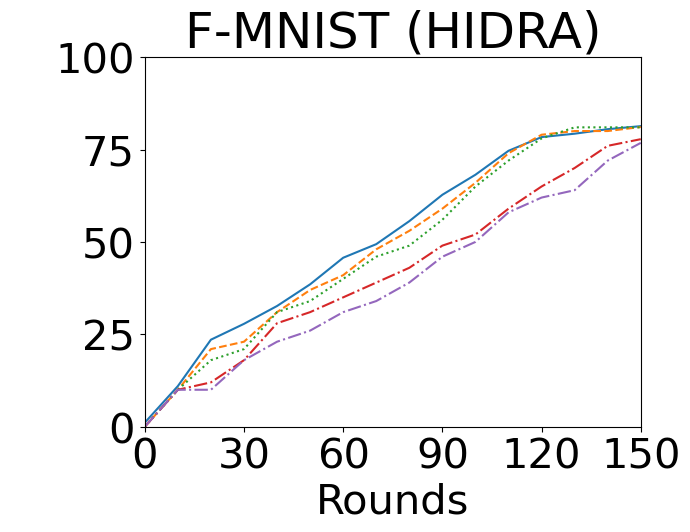} 
    \caption{Model accuracy over training steps 
    when \randeigen is applied to models trained on
    MNIST and F-MNIST datasets, under different choices of $\epsilon$, remains high.}
    \label{fig:img_eval_diff_eps}
\end{figure}

\begin{figure}
    \centering
    \includegraphics[width=0.49\linewidth]{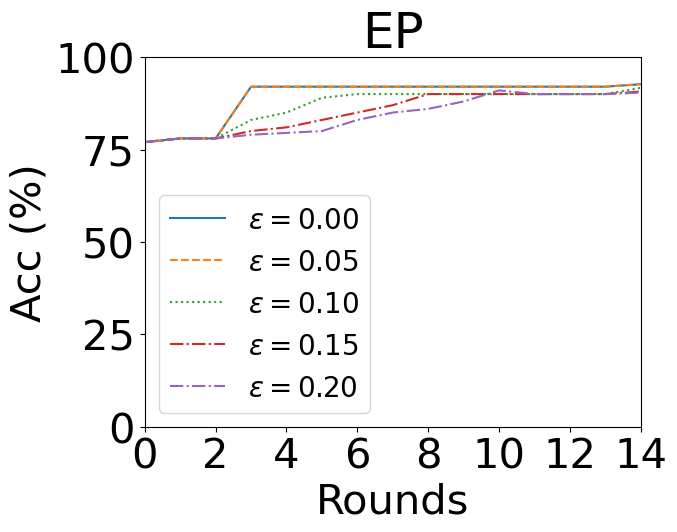}
    \includegraphics[width=0.49\linewidth]{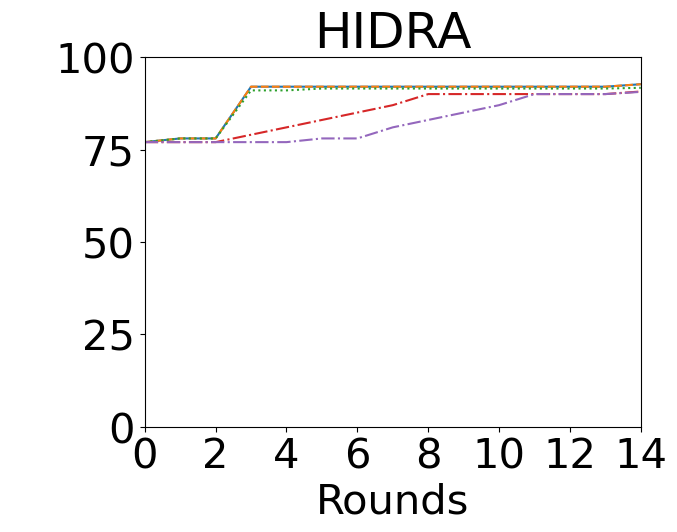}
    \caption{Model accuracy over training steps when the \randeigen~defense is applied to BERT language model trained on SST dataset, under different choices of $\epsilon$, remains high.}
    \label{fig:acc_sst}
\end{figure}
Though our theoretical analysis is for $\epsilon < 1/12$, we evaluate \randeigen~with different $\epsilon \in \{0.00, 0.05, 0.10, 0.15, 0.20\}$ with different attacks on respective datasets. The accuracy across all the training steps for all $\epsilon$ values is illustrated in Figure \ref{fig:img_eval_diff_eps} and \ref{fig:acc_sst}. The accuracy remains high, showing that \randeigen~is effective in mitigating all these attacks across the range of $\epsilon$ tested. 
We have also evaluated \randeigen~ against $\epsilon > 0.2$, and observed that it achieves similar results up to $\epsilon = 0.33$. Beyond that, the model accuracy either drops significantly, or model accuracy is maintained but ASR is as high as without defenses.
We delve deeper on the configuration with largest adversarial corruption $\epsilon=0.2$.

The model accuracy with \randeigen~deployed against attacks at $\epsilon=0.2$, for image classification and language tasks in  Tables \ref{table:baseline_img} and \ref{table:baseline_sst} respectively. Overall, 
we observe that \randeigen~effectively mitigates the {\em all of aforementioned attacks across all evaluated configurations}. Notably, \randeigen~also filters out poisoned samples generated by the {\em adaptive} attack strategy of HiDRA against prior strong robust aggregators including  FILTERING and NO-REGRET. This is because of two reasons:
(1) \randeigen~does not need to use any chunking like in prior defenses; and (2) it uses a convergence criterion rather than a sharp threshold. HiDRA overfits its attack strategy to the latter threshold to attack the deterministic defenses.

For example in Table \ref{table:baseline_img}, \randeigen~retains the original accuracy of the model during training even in the presence of the HiDRA poisoning attacks after training, whereas all baselines evaluated experience a drastic drop in accuracy (over $80\%$).

We experimentally also observe that \randeigen~outperforms prior works on FILTERING and NO-REGRET, both in terms of model 
accuracy and reduced ASR, in some of the non-adaptive attacks though the difference is marginal in most cases.  

\begin{mdframed}[backgroundcolor=mygray]
    \randeigen~achieves comparable performance to existing practical implementations of
    strong robust aggregators on non-adaptive attacks, and further fully mitigates the adaptive HiDRA attack designed to circumvent robust aggregators.
\end{mdframed}

\subsection{Extension to the Centralized Setup} \label{subsection:eval_centralized}

Our primary evaluation has been on attacks in the federated setup where the adversary has direct control over the gradients. Nonetheless, \randeigen~makes no assumptions specific to the federated setup, and 
can be used during centralized training.

To confirm this, we show $2$ attacks in the centralized setup:
Gradient matching \cite{gradient_matching} and subpopulation attack \cite{subpopulation}. These are representative of state-of-the-art data poisoning in the centralized setup which achieve a high ASR against prior defenses, with a small $\epsilon$ of below $0.05$. We used the default parameters and models recommended in their respective papers on the MNIST dataset. These are targeted poisoning attacks, designed to craft poisoned datasets that cause the model to misclassify inputs from specific target classes when used for training. 
The results of our evaluations are summarized in Table \ref{tab:additional_expt}.
We see that \randeigen~causes the attack success rate to drop from over $85\%$ without the defense to below $5\%$. At the same time, the model accuracy drop is less than $1.5\%$. It shows that \randeigen~is highly effective and these results are broadly consistent with those observed for the federated setup. This occurs because the gradients computed from poisoned samples are statistical outliers relative to those from clean samples (see Figure \ref{subfloat:a_gm} and \ref{subfloat:b_sp}), enabling \randeigen~to filter them out. 

\begin{table}[]
\centering
\caption{Model Accuracy (ACC) and the Attack Success Rate (ASR) of the gradient matching and subpopulation attack. }
\label{tab:additional_expt}
\begin{tabular}{|l|l|l|l|}
\hline
Attack & Defense    & ACC    & ASR   \\ \hline
\multirow{2}{*}{\begin{tabular}[c]{@{}l@{}}Gradient\\ Matching\end{tabular}}    & No Defense & 94.0\% & 87.0\% \\ \cline{2-4} 
       & \randeigen~& 93.1\% & 4.2\% \\ \hline
\multirow{2}{*}{\begin{tabular}[c]{@{}l@{}}Subpopulation\\ Attack\end{tabular}} & No Defense & 92.7\% & 93.0\% \\ \cline{2-4} 
       & \randeigen~& 92.5\% & 3.1\% \\ \hline
\end{tabular}
\end{table}

\subsection{Post-training attacks Adapted to In-training} \label{subsection:dfba}

We have shown the effectiveness of \randeigen~for filtering out outlier gradient vectors during training. There are other post-training attacks, such as the Data Free Backdoor Attack (DFBA)~\cite{dfba}, which edit the model parameters after training. These are
technically are not within scope of any defense deployed during training. However, one could wonder if such attacks can be adapted to work during training and whether \randeigen~would be effective if so.

In order to evaluate \randeigen~for this case, we adapted DFBA for the federated learning, as follows:
\begin{enumerate}
    \item At the beginning of each training step, a shared centralized model is distributed to all participants for local training.
    \item After local training, each participant sends their computed model gradients to the centralized server, which aggregates them to update the global model.
    \item Malicious participants apply the DFBA attack to the received model to generate a backdoored version, at each training step, and then return the gradients derived from this backdoored model to the server.
\end{enumerate}

\begin{figure*}
    \centering
    \subfloat[ \label{subfloat:a_gm}]{\includegraphics[width=0.33\linewidth]{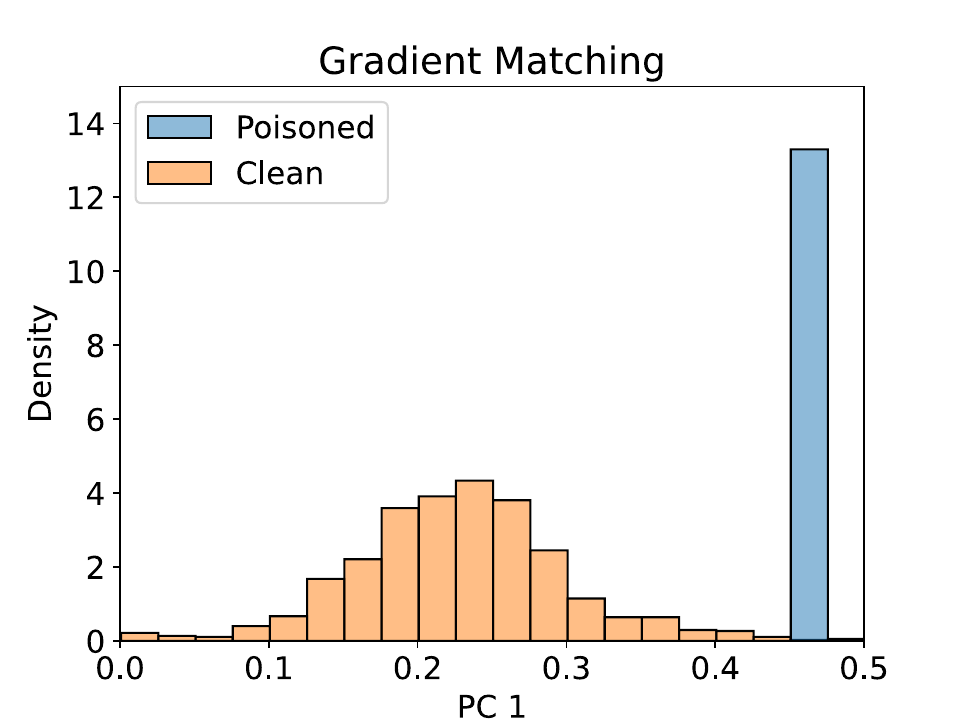}} 
    \subfloat[\label{subfloat:b_sp}]{\includegraphics[width=0.33\linewidth]{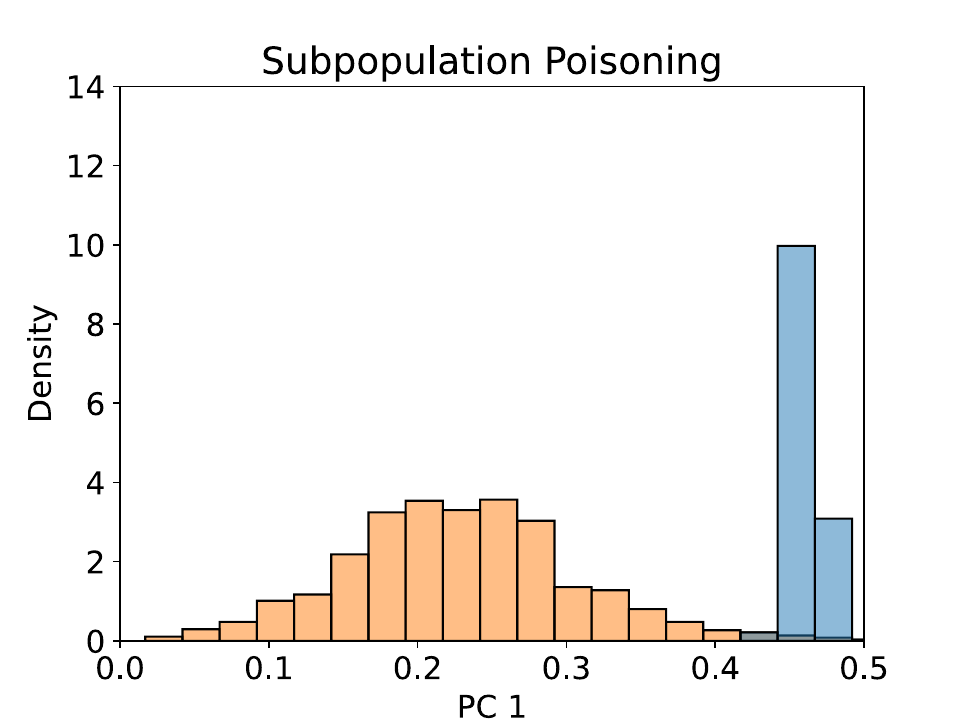}} 
    \subfloat[\label{subfloat:c_dfba}]{\includegraphics[width=0.33\linewidth]{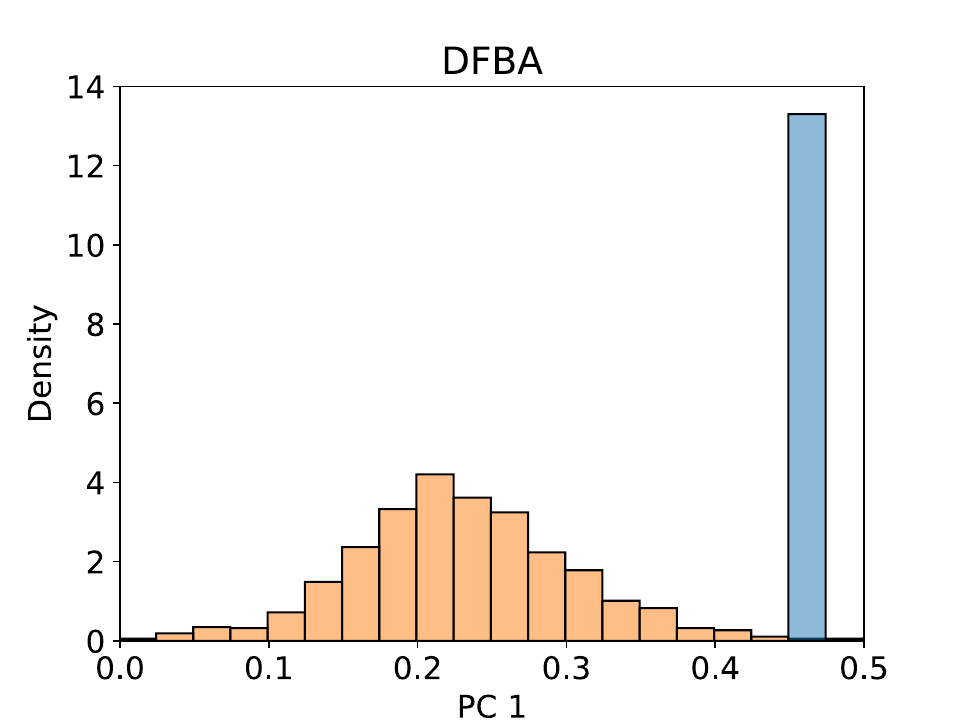}} 
    \caption{Distribution of the poisoned and clean gradients projected against the dominant eigenvector (PC 1) for the Gradient Matching, Subpopulation Matching and DFBA attacks in Figures \ref{subfloat:a_gm}, \ref{subfloat:b_sp} and \ref{subfloat:c_dfba}, respectively. The gradients shown in Figures \ref{subfloat:a_gm} and \ref{subfloat:b_sp} were sampled from the first training step, whereas those in Figure \ref{subfloat:c_dfba} were from a randomly sampled training step.}
    \label{fig:gradients_others}
\end{figure*}

The results of running \randeigen~in this adapted DBFA on the MNIST dataset in Table \ref{tab:additional_expt_dbfa}. Consistent with the other experiments in Section \ref{section:expt}, \randeigen~successfully nullifies the DFBA attack adapted for the federated setting. The ASR drops from over $94\%$ to below $2.5\%$, while the model accuracy does not drop by more than $1.5\%$ as a result of filtering with~\randeigen. The gradients from the poisoned samples  are statistical outliers relative to the clean ones (see Figure \ref{subfloat:c_dfba}), enabling \randeigen~to filter them out.

\begin{table}[]
\centering
\caption{Model Accuracy (ACC) and Attack Success Rate (ASR) of the DFBA attack, evaluated under varying $\epsilon$.}
\label{tab:additional_expt_dbfa}
\begin{tabular}{|l|l|l|l|}
\hline
$\epsilon$            & Defense    & ACC    & ASR    \\ \hline
\multirow{2}{*}{0.10} & No Defense & 94.6\% & 94.0\% \\ \cline{2-4} 
                      & \randeigen~& 94.3\% & 1.9\%  \\ \hline
\multirow{2}{*}{0.15} & No Defense & 94.6\% & 94.3\% \\ \cline{2-4} 
                      & \randeigen~& 94.3\% & 2.5\%  \\ \hline
\multirow{2}{*}{0.20} & No Defense & 93.0\% & 94.3\% \\ \cline{2-4} 
                      & \randeigen~& 92.4\% & 2.4\%  \\ \hline
\end{tabular}
\vspace{-5mm}
\end{table}

\section{Related Works}

\paragraph{Poisoning Attacks}
Poisoning attacks involve manipulating a subset of the samples to achieve
some adversarial goal (e.g. manipulating model behavior over certain types of inputs
or reducing overall model accuracy).
There are three main types of poisoning attacks \cite{DBLP:journals/ieeesp/JereFK21}: backdoor, targeted, 
and untargeted attacks. Backdoor attacks involve inserting a malicious
functionality into the model, for example, outputting a particular result when a
trigger is present. On the other hand, targeted attacks degrade a model's 
performance on specific data subpopulations (e.g. particular classes in image classification). Finally, untargeted attacks involve 
aim to indiscriminately 
degrade the overall accuracy of a FL model. 
All three types of attacks can work in the federated learning
setting~\cite{zhu2023byzantine, dnc, DBLP:journals/ieeesp/JereFK21, DBLP:journals/access/XiaCYM23}. One of the earliest works on poisoning attacks for deep learning systems used targeted poisoning on federated learning setups~\cite{shen2016auror}. 

Many prior works have independently observed that poisoned samples deviate significantly from the distribution
over the clean samples~\cite{shen2016auror,dnc,li20233dfed}, which motivates norm-bounding as a basis for a generic defense against
data poisoning attacks. There has been recent work to develop data poisoning attacks which bypass
detection. 
For instance, 
Drupe \cite{DBLP:conf/sp/TaoWFSMZ24} is data poisoning attack which is trying to construct poisoned samples that have embeddings (not gradients) statistically indistinguishable from the distribution of those for clean samples. If the same is achievable for gradients, norm-bounding defenses would not defeat such an attack.

\paragraph {Byzantine Robust Aggregators}
Existing literature (e.g. \cite{DBLP:conf/sp/ChoudharyKS24})
characterizes Byzantine robust aggregators as strong or weak, depending on whether they achieve optimal information-theoretic bias bounds.
Examples of recent Byzantine robust aggregators, together with their computational complexity and bias bounds, are in Table~\ref{table:priorwk}.

Weak robust aggregators admit a worst-case upper bound of $O(\sqrt{\epsilon d})$
on the maximum bias incurred. These aggregators typically make use of dimension-wise locality statistics
to mitigate the presence of corrupted data
\cite{pillutla2022robust, DBLP:conf/emnlp/Zhang0022a,
DBLP:journals/corr/abs-1802-10116}. Fundamentally, any
statistic computed locally dimension-wise incurs a bias proportional to $O(\sqrt{\epsilon d})$, 
as small bias compounds across dimensions. The bias can be too high for high-dimensional settings. 
Examples of weak robust aggregators
include taking the median across the samples \cite{yin2018byzantine}, 
trimmed means \cite{yin2018byzantine},
and geometric median \cite{pillutla2022robust, blanchard2017machine, chen2020distributed}.
Weak robust aggregators are usually designed with specific
attacks or assumptions in mind, and have been shown to be circumvented by other
attacks (see for example,
\cite{xie2019dba, xie2020fall, fang2020local}).


Strong robust aggregators identify possible outliers based on their projection on the dominant eigenvector of the covariance of samples~\cite{diakonikolas2023algorithmic}.  
The work of~\cite{zhu2023byzantine} demonstrated 
that practical implementations of strong robust aggregators effectively mitigate many non-adaptive (prior) attacks 
aimed at introducing significant bias into the aggregates within the federated learning framework.
However, such aggregators require $\Omega(\epsilon n d^2)$ operations and $\Omega(d^2)$ memory,
making them impractical when $d$ is large (e.g. ResNet 18 has approx. $2^{23}$
parameters). The practical inefficiency that is inherent when dealing with real-world poisoning attacks is known, see for example the work on DnC~\cite{dnc}. The dimensionality reduction used therein does {\em not} yield a strong robust aggregator, however. Although DnC \cite{dnc} admits a similar filtering-based approach to mitigate poisoning attacks has known attacks, 
we note that it fails to limit bias against adaptive attacks, 
as it only considers poisoned data which aligns with the direction of the dominant eigenvector
(see~\cite{DBLP:conf/sp/ChoudharyKS24}, Appendix B for a more detailed discussion).
More recently, HIDRA has been introduced as an 
untargeted poisoning attack designed to attack 
practical implementations of strong robust aggregators, 
resulting in a bias proportional to $\sqrt{d}$. Their strategy is provably near-optimal, though its specific implementation we evaluate is known to be not stealthy---it overfits to sharp thresholds used by the defense, and hence our defense is able to nullify it.

Finally, while our focus is on poisoning attacks, filtering outliers to shape the gradient average may have broader use in differentially-private training~\cite{dpsgd} or mitigating inference attacks~\cite{xu2024robust}.



\section{Conclusions}

We have presented the first robust aggregator with strong bias bounds and with quasi-linear running time. It is usable practically for high dimensional distributions arising in modern neural network training. As a generic norm-bounding defense, we experimentally find that it nullifies many existing poisoning attacks and has marginal loss in training accuracy in absence of attacks.

\begin{acks}
We thank the anonymous reviewers for their feedback. We also thank Mallika Prabhakar and artefact evaluation committee members for testing the artifacts, and Sarthak Choudhary for valuable comments on the manuscript. This research is supported by a Singapore Ministry of Education (MOE) Tier 2 grant MOE-T2EP20124-0007 and CRYSTAL Centre, National University of Singapore.
\end{acks}
\bibliographystyle{ACM-Reference-Format}
\bibliography{ref.bib}

\appendix

\section{Proof of Theorem \ref{thm:bias_revised}} 
\label{appendix:thm}

Let  $S$ be the set of clean samples, and assume that is a $(5\epsilon, \delta)$-stable set (as per Definition \ref{dfn:epsilon_delta_stable}). Construct the $\epsilon$-corrupted set $X$ by corrupting an $\epsilon$-proportion of samples from $S$. Denote $X_i$ to be the samples remaining after the $i$-th iteration of filtering, for $i=1, \dots, \tau$, where $\tau \leq 2n\epsilon$ is the number of iterations taken by Algorithm \ref{alg:filter_modified}.
Correspondingly, let $S_i = S \cap X_i$ and $T_i = X_i \setminus S_i$ denote the clean and corrupted samples remaining at the start of the $i$-th iteration, respectively. Let $\mu_A$ and $\Sigma_A$ denote the empirical mean and covariance matrix of a set of samples $A$. After $\tau$ filtering iterations,  the robust aggregate of $X_{\tau+1}$ is returned in Line 15 of Algorithm \ref{alg:filter_modified}.  Henceforth, we will use $\cdot$ to represent dot product, instead of $\langle \cdot, \cdot \rangle$.

In Lemma \ref{lemma:balancing}, we prove that for a set $X_i$ that can be partitioned into subsets $S_i$ and $T_i$, the sum of deviations between the points in $S_i$ and the mean of $X_i$  must equal the corresponding sum for $T_i$. 
\begin{lemma}[\emph{Balancing Lemma}] \label{lemma:balancing}
    For any $i \leq \tau$, let $w_i = \frac{|S_i|}{|T_i|}$ and the set $X_i = S_i \cup T_i$. 
    Then, for any unit vector $v$, the following holds:
    \[
        w_i v\cdot( \mu_{S_i} - \mu_{X_i}) = - v\cdot( \mu_{T_i} - \mu_{X_i})
    \]
\end{lemma}
\begin{proof}
    From the definition of the sample mean of $X_i$,
    \[ 0 = v\cdot \left(\left(\sum_{x\in X_i}{x_i}\right) - n\mu_{X_i} \right)= v\dot\sum_{x\in X_i}(x_i - \mu_{X_i}).  \]
    Therefore, 
   \[
    \sum_{x \in X_i} v\cdot(x - \mu_{X_i}) = \sum_{s \in S_i} v\cdot(s - \mu_{X_i}) + \sum_{t \in T_i} v\cdot(t - \mu_{X_i}), 
    \]
    or equivalently, 
    \[
        \sum_{s \in S_i} v\cdot(s - \mu_{X_i}) = - \sum_{t \in T_i} v\cdot(t - \mu_{X_i}).
    \]
    Since $\sum_{s \in S_i} v\cdot(s - \mu_{X_i}) = |S_i| v\cdot( \mu_{S_i} - \mu_{X_i})$ and $ \sum_{t \in T_i} v\cdot(t - \mu_{X_i}) = |T_i| v\cdot( \mu_{T_i} - \mu_{X_i}) $, we have
    \[
        |S_i| v\cdot( \mu_{S_i} - \mu_{X_i}) = -|T_i| v\cdot( \mu_{T_i} - \mu_{X_i})
    \]
    Dividing both sides by $|T_i|$, 
    \[
        w_i v\cdot( \mu_{S_i} - \mu_{X_i}) =  -v\cdot( \mu_{T_i} - \mu_{X_i}).
    \]
\end{proof}

In Lemma \ref{lemma:balancing_epsilon} and \ref{lemma:max_deviation_mu_s}, we establish an upper bound for the proportion of samples from $S$ which deviate from $\mu_S$ when projected along any unit vector by a factor of $\frac{\delta}{\epsilon}$. Additionally, we analyze the effect of including such deviating samples in the computation of the sample mean.
\begin{lemma} \label{lemma:balancing_epsilon}
    Let $S$ be a $(5\epsilon, \delta)$-stable set for some constant $\epsilon \leq \frac{1}{12}$. There does not exist a set $S_\epsilon \subset S$ where $|S_\epsilon| = \epsilon |S|$, and for any unit vector $v$ , $|v\cdot(\mu_{S_\epsilon} - \mu_S)| \geq \frac{\delta}{\epsilon}$ .
\end{lemma}
\begin{proof}
    Suppose such a $S_\epsilon$ exists. Let $S_\epsilon' = S \setminus S_\epsilon$. By Balancing Lemma (Lemma \ref{lemma:balancing}), set $X_i=S$, $T_i = S_\epsilon$ and $S_i = S_\epsilon'$, and we obtain
    \[
        \frac{|S_\epsilon'|}{|S_\epsilon|} v\cdot( \mu_{S_\epsilon'} - \mu_{S}) =  -v\cdot( \mu_{S_\epsilon} - \mu_{S}).
    \]
    Since $|v\cdot(\mu_{S_\epsilon} - \mu_S)| \geq \frac{\delta}{|S_\epsilon|/|S|}$, 
    \[
        \frac{|S_\epsilon'|}{|S_\epsilon|} |v\cdot( \mu_{S_\epsilon'} - \mu_{S})| \geq\frac{\delta}{|S_\epsilon|/|S|},
    \]
    or equivalently, 
    \[ 
        |v\cdot( \mu_{S_\epsilon'} - \mu_{S})| \geq  \frac{|S_\epsilon|}{|S_\epsilon'|}  \frac{\delta}{|S_\epsilon|/|S|}  = \frac{|S|}{|S_\epsilon'|}\delta > \delta,
    \]
    which violates the $(5\epsilon, \delta)$-stablity of $S$ since $|S_\epsilon'| \geq (1-5\epsilon)|S|$. Thus, by contradiction, such $S_\epsilon$ does not exist. 
\end{proof}

In Lemma \ref{lemma:max_deviation_mu_s} and Corollary \ref{corr:max_deviation}, we analyze the natural variance of samples in $S$. In particular, in Lemma \ref{lemma:max_deviation_mu_s}, we show that no more than an $\epsilon$-proportion of samples from $S$ deviate from the population mean $\mu_S$ by more than $2\frac{\delta}{\epsilon}$ when projected along any unit vector. Consequently, in Corollary \ref{corr:max_deviation}, we show that this deviation is at most $2\frac{\delta}{2} + \delta$ when we consider the sample mean $\mu_{\tilde{S}}$, where $\tilde{S}$ is any subset of $S$ with $|\tilde{S}|\geq (1-5\epsilon)n$.

\begin{lemma} \label{lemma:max_deviation_mu_s}
    Let $S$ be a $(5\epsilon, \delta)$-stable set for some constant $\epsilon \leq \frac{1}{12}$. Let $S' = \left\{s \in S : |v \cdot (s - \mu_S)| > 2\frac{\delta}{\epsilon} \text{ for all unit vector $v$}\right\}$. Then $|S'| \leq \epsilon |S|$
\end{lemma}
\begin{proof}
    
    We prove this by contradiction; suppose that $|S'| > \epsilon |S|$, and $S'' \subseteq S'$ is any subset of $S'$ where $|S''| = \epsilon|S|$. Partition $S'' = S_\alpha \cup S_\beta$ where $S_\alpha = \{s \in S'': v\cdot(s - \mu_S) < 0\}$ and $S_\beta = S'' \setminus S_\alpha$. Without loss of generality, assume that $|S_\alpha|  \geq |S_\beta|$. Thus, $\epsilon|S| \geq |S_\alpha| \geq \frac{|S''|}{2} = |S| \left(\frac{\epsilon}{2}\right) \geq |S_\beta|$, and 
    \[
    \begin{split}
       |v \cdot (\mu_{S_\alpha} - \mu_S)| &= \frac{1}{|S_\alpha|}\left|\sum_{s \in S_\alpha} v\cdot(s - \mu_S) \right| \\
       & \stackrel{1}{=} \frac{1}{|S_\alpha|}\sum_{s \in S_\alpha} |v\cdot(s - \mu_S) | \\ 
       & \stackrel{2}{\geq} \frac{1}{|S| \frac{\epsilon}{2}} \sum_{s \in S_\alpha} |v\cdot(s - \mu_S) | \\
       &\stackrel{3}{>} \frac{1}{|S| \frac{\epsilon}{2}}  |S| \frac{\epsilon}{2} \frac{2\delta}{\epsilon} \\
       &= 2\frac{\delta}{\epsilon}
    \end{split},
    \]
    where $\stackrel{1}{=}$ is due to the definition of $S_\alpha = \{s \in S'': v\cdot(s - \mu_S) < 0\}$,  $\stackrel{2}{=}$ is due to $|S_\alpha| \geq |S| \frac{\epsilon}{2}$, and $\stackrel{3}{>}$ is due to the definition of $S'$. However, $|v \cdot (\mu_{S_\alpha} - \mu_S)| > 2\frac{\delta}{\epsilon}$ contradicts Lemma \ref{lemma:balancing_epsilon}. The case where $|S_\beta| > |S_\alpha|$ follows the same argument. Thus, $|S'| \leq \epsilon |S|$
\end{proof}
    
\begin{corollary} \label{corr:max_deviation}
    Continuing from Lemma \ref{lemma:max_deviation_mu_s}, let $$S'' = \left\{s \in S : |v \cdot (s - \mu_{\tilde{S}})| > 2\frac{\delta}{\epsilon} + \delta \text{ for all unit vector $v$}\right\}$$ where $\tilde{S}$ is any set where $\tilde{S} \subseteq S$ and $|\tilde{S}| \geq (1-5\epsilon)|S|$.Then $|S''| \leq \epsilon |S|$
\end{corollary}
\begin{proof}
    By Definition \ref{dfn:epsilon_delta_stable} (1), $|v \cdot (\mu_{S} - \mu_{S_i})| \leq \delta$, and consider $s \in S$ where $|v \cdot (s - \mu_{\tilde{S}})| > 2\frac{\delta}{\epsilon} + \delta$. Thus, 
    \[
    \begin{split}
        2\frac{\delta}{\epsilon} + \delta < |v \cdot (s - \mu_{\tilde{S}} )| &= |v \cdot (s - \mu_{S}) + v \cdot (\mu_{S} - \mu_{\tilde{S}})| \\
        & \leq  |v \cdot (s - \mu_{S})| + | v \cdot (\mu_{S} - \mu_{\tilde{S}})| \\
        & \leq |v \cdot (s - \mu_{S})| + \delta.
    \end{split}
    \]
    where the final inequality is due to the Triangle inequality. 
    Note that  $|v \cdot (s - \mu_{\tilde{S}})| > 2\frac{\delta}{\epsilon} + \delta \Rightarrow|v \cdot (s - \mu_{S})| > 2 \frac{\delta}{\epsilon}$. 
    Thus, $S'' \subseteq S'$, where $S'$ is as defined in Lemma \ref{lemma:max_deviation_mu_s}, and hence, $|S''| \leq |S'| \leq \epsilon |S|$.
\end{proof}

With Corollary \ref{corr:max_deviation}, we formally define ``inlier'' and ``outlier'' samples of an $\epsilon$-corrupted version of $S$ in Definition \ref{definition:outlier}. 
\begin{definition} \label{definition:outlier}
    Let $S$ be a $(5\epsilon, \delta)$-stable set, and $X$ is an $\epsilon$-corrupted version of $S$, and $\delta' = \delta + 2\frac{\delta}{\epsilon}$ Then, $X$ can be partitioned into two disjoint sets $X=S'\cup T'$, where
    \begin{equation*}
    \begin{split}
        S' = \{ x \in X&: v \cdot(x - \mu_{\tilde{S}}) \leq \delta'  \\
        & \quad \text{ for all unit vectors $v$, all $\tilde{S} \subseteq S$ s.t.  $|\tilde{S}|\geq(1-5\epsilon)|S|$}\},
    \end{split}
    \end{equation*}
    denotes the set of \textbf{inliers}, and $ T' = X \setminus S'$,
    denotes the set of \textbf{outliers}.
\end{definition}

Corollary~\ref{corr:max_deviation} has shown that at most $\epsilon|S| \leq n\epsilon$ outliers can exist in $S$. Furthermore, $T$ can introduce at most $n\epsilon$ more, so the total number of outliers in any $X_i \subseteq X$ is at most $2n\epsilon$.

In Lemma \ref{lemma:expected_removal}, we prove that the expected number of inlier samples removed in each iteration of Algorithm \ref{alg:filter_modified} is upper bounded by the expected number of outlier samples removed. 
\begin{lemma} \label{lemma:expected_removal}
    Let $\delta' =  2\frac{\delta}{\epsilon} + \delta$, and define $S'_i = X_i \cap S'$ and $T_i' = X_i \setminus S_i'$, where the set of inlier samples $S'$ is defined in Definition \ref{definition:outlier}. 
    
    For any iteration $i < \tau$, if $|v\cdot(\mu_{S_i'} - \mu_{X_i})|  > \delta'$, where $v$ is dominant eigenvector computed in iteration $i$, the following holds: $$\IE[|S_i'| - |S_{i+1}'| | X_i] \leq \IE[|T_i'| - |T_{i+1}'| | X_i] $$ That is, the expected number of samples removed from $S_i'$ is upper bounded by the expected number of samples removed from $T_i'$. 
\end{lemma}
\begin{proof}
    Without loss of generality, assume that $v\cdot(\mu_{S_i} - \mu_{X_i}) < 0$.  Since  $v\cdot(\mu_{S_i} - \mu_{X_i}) < -\delta'$, then by the construction of $S_i'$, for all $s \in S_i'$, $v\cdot(s - \mu_{X_i}) < 0$. 
    Thus, 
    \begin{equation} \label{eqn:balanced}
        \frac{1}{|S_i'|} \left|\sum_{s \in S_i'} v\cdot(s - \mu_{X_i}) \right| = \frac{1}{|S_i'|} \sum_{s \in S_i'} |v\cdot(s - \mu_{X_i})|. 
    \end{equation}
    Furthermore,
    \[
    \begin{split}
        |v \cdot(\mu_{T_i'} - \mu_{X_i})| & \stackrel{1}{=} \frac{|S_i'|}{|T_i'|}|v \cdot(\mu_{S_i'} - \mu_{X_i})| \\
        & = \frac{|S_i'|}{|T_i'|}\frac{1}{|S_i'|} \left|\sum_{s \in S_i'} v\cdot(s - \mu_{X_i}) \right| \\
        & \stackrel{2}{=}  \frac{|S_i'|}{|T_i'|}\frac{1}{|S_i'|} \sum_{s \in S_i'} |v\cdot(s - \mu_{X_i})|  \\
        & = \frac{1}{|T_i'|}\sum_{s \in S_i'} |v\cdot(s - \mu_{X_i})|  
    \end{split}
    \]
    where $\stackrel{1}{=}$ and $\stackrel{2}{=}$ are due to the Balance Lemma (Lemma \ref{lemma:balancing}) and (\ref{eqn:balanced}), respectively. Thus, 
    \[
    \frac{1}{|T_i'|} \sum_{t \in T_i'} |v\cdot(t - \mu_{X_i})| \geq |v \cdot(\mu_{T_i'} - \mu_{X_i})| = \frac{1}{|T_i'|} \sum_{s \in S_i} |v\cdot(s - \mu_{X_i}) |.
    \]
    Multiplying by $|T_i|$,
    \[
     \sum_{t \in T_i'} |v\cdot(t - \mu_{X_i})| \geq  \sum_{s \in S_i'} |v\cdot(s - \mu_{X_i}) |.
    \]
    To bound the expectations, first note that  each point $x \in X_i$ is removed with probability $\frac{|v\cdot(x - \mu_{X_i})|}{\max \mathcal{P}}$ independently, where $\max \mathcal{P}$ is as defined in Line 12 of Algorithm \ref{alg:filter_modified}. Thus, 
    \[
    \begin{split}
     \IE[|T_i'| - |T_{i+1'}| | X_i]  &= \frac{1}{\max \mathcal{P}}\sum_{t \in T_i'} |v\cdot(t - \mu_{X_i})| \\
     & \geq  \frac{1}{\max \mathcal{P}}\sum_{s \in S_i'} |v\cdot(s - \mu_{X_i'}) | \\
     & = \IE[(|S_i'| - |S_{i+1}'|)| X_i] .
    \end{split}
    \]
\end{proof}

On expectation, the total number of inlier samples removed  across all rounds of Algorithm \ref{alg:filter_modified} is upper bounded by the total number of outlier samples removed, by linearity of expectations applied on Lemma \ref{lemma:expected_removal}. In fact, the following corollary bounds the difference in the total inliers and outliers removed with high probability. 

\begin{corollary} \label{corr:whp_difference}
    Let $N_S = |S'_1| - |S'_{\tau+1}|$ be the number of inliers removed across all $\tau$ iterations, and $N_T = |T'_1| - |T'_{\tau+1}|$ similarly. If $|T_\tau| > 0$, the probability $\IP[N_S - N_T \geq n\epsilon] \leq \exp(-\Omega(n\epsilon))$. 
\end{corollary}
We omit the proof of Corollary \ref{corr:whp_difference} here, as it is the same as \cite{diakonikolas2023algorithmic}, Exercise 2.11. 



Next, we proceed to analyze the correctness of the eigenvalue convergence check (Line 3) of Algorithm \ref{alg:filter_modified}. In Lemma \ref{lem:martingale}, we show that the proportion of outlier samples remaining in each iteration decreases on expectation. 


\begin{lemma}\label{lem:martingale}
For all $i < \tau$, let $\epsilon_i' =\frac{|T_i'|}{|X_i'|}$, where $X_i = S_i' \cup T_i'$ and $S_i', T_i'$ are as defined in Lemma \ref{lemma:expected_removal} and its conditions hold. Then, $ \IE\left[\epsilon'_{i+1} \middle | X_i \right] \leq \epsilon_i'$. Thus, the sequence $\epsilon'_{1}, \dots, \epsilon_\tau'$ is a supermartingale.
\end{lemma}

\begin{proof}
    For a fixed iteration $i$, define the random variables $N_{S_i}:= |S_i'| - |S_{i+1}'| | X_i$ and $N_{T_i} :=|T_i'| - |T_{i+1}'| | X_i$. Thus, 
    \[
    \begin{split}
        \IE\left[\frac{|T_{i+1}'|}{|S_{i+1}'|} \middle | X_i \right] &= \IE\left[\frac{|T_{i}'| - N_{T_i}}{|S_{i}'| - N_{S_i}} \middle | X_i \right] \\
        & \stackrel{1}{=} \frac{\IE[|T_{i}'| - N_{T_i}|X_i]}{\IE[|S_{i}'| - N_{S_i}|X_i]} \\
        & \stackrel{2}{\leq} \frac{\IE[|T_{i}'| - N_{T_i}|X_i]}{\IE[|S_{i}'| - N_{T_i}|X_i]} \\
        & \leq \frac{|T_i'|}{|S'_i|}.
    \end{split}
    \]
    The last equality is because $N_{S_i}|X_i$ and  $N_{T_i}|X_i$ are independent. The first inequality is because $\IE[N_{S_i}] \leq \IE[N_{T_i}]$ (Lemma~\ref{lemma:expected_removal}). Thus,
    \[
        \IE[(1-\epsilon_{i+1}')|X_i] = \left(1+\IE\left[\frac{|T_{i+1}'|}{|S_{i+1}'|} \middle | X_i \right] \right)^{-1} \geq \left( 1+\frac{|T_i'|}{|S'_i|}\right)^{-1} = 1-\epsilon_i',
    \]
    and so $\IE[\epsilon_{i+1}'|X_i] \leq \epsilon'_i$.
    By definition $\forall{j},\epsilon'_{j} \in [0, 1]$, hence $\IE[|\epsilon'_{j}|] < \infty$. 
    Therefore, $\epsilon'_{1}, \dots, \epsilon_\tau'$ is a supermartingale sequence. 
\end{proof}

\begin{lemma}\label{lemma:sigma_xi_decreasing}
Let $\lambda_i := ||\Sigma_{X_i}||_2$ denote the dominant eigenvalue computed during the $i$-th iteration of Algorithm \ref{alg:filter_modified}, and the conditions in Lemma \ref{lem:martingale} holds.  Then, the sequence $\IE[||\Sigma_{X_i}||_2 - ||\Sigma_{S_i'}||_2]$ converges to 0, for the sequence $\epsilon'_1, \dots, \epsilon'_\tau$ in Lemma \ref{lem:martingale}. In other words, the eigenvalue sequence $\lambda_1, \dots, \lambda_\tau$ converges to $||\Sigma_{S_i'}||_2$ on expectation. 
\end{lemma}

\begin{proof}
        The following is an algebraic fact. For arbitrary subsets $S'_i$ and $T'_i$, where $S'_i \cup T'_i = X_i$:
    \begin{equation} 
        \Sigma_{X_i} = (1-\epsilon'_i) \Sigma_{S_i'} + \epsilon'_i \Sigma_{T_i'} + \epsilon'_i (1-\epsilon'_i) (\mu_{S_i'} - \mu_{T_i'})(\mu_{S_i'} - \mu_{T_i'})^T,
    \end{equation}
    This can be rewritten as:
    \begin{equation} \label{eqn:sigma_xi_2}
        \Sigma_{X_i} - \Sigma_{S_i'} = \epsilon'_i (\Sigma_{T_i'} -  \Sigma_{S_i'}) + \epsilon'_i (1-\epsilon'_i) (\mu_{S_i'} - \mu_{T_i'})(\mu_{S_i'} - \mu_{T_i'})^T,
    \end{equation}
    
    By Lemma \ref{lem:martingale}, $\epsilon'_{1}, \dots, \epsilon_\tau'$ is a supermartingale sequence. Thus, the terms in (\ref{eqn:sigma_xi_2}) which are all proportional to $\epsilon_i'$ are decreasing and converging to 0
    in expectation. Thus, 
    $\Sigma_{X_i}$ is converging to $\Sigma_{S_i'}$. Hence, the sequence $\IE[||\Sigma_{X_i}||_2 - ||\Sigma_{S_i'}||_2]$  is decreasing and converges to 0, or equivalently, $\IE[\lambda_i] =\IE[||\Sigma_{X_i}||_2] \rightarrow \IE[||\Sigma_{S_i'}||_2]$. 
\end{proof}

Thus far we have shown that the eigenvalue in each iteration is converging towards $||\Sigma_{S'_i}||_2 $ on expectation. We show in Lemma~\ref{lemma:lambda_decrease} that on expectation the eigenvalue is decreasing in each successive iteration, using Lemma~\ref{lemma:helper} as an intermediate step.

\begin{lemma} [\emph{Helper Lemma}] \label{lemma:helper}
Let $R_i \subseteq X_i$ denote the samples removed in the $i$-th round of Algorithm \ref{alg:filter_modified}. Then, we have
\[
    \IE\left[\sum_{r \in R_i} (v_i \cdot(r - \mu_{X_i}))^2\right] 
    \leq \lambda_i \IE[|R_i|],
\]
where $\lambda_i$ and $v_i$ denote the dominant eigenvalue and eigenvector, respectively.


\end{lemma}
\begin{proof}
    A sample $x \in X_i$ is removed with probability $p_x = \frac{|v_i \cdot(x - \mu_{X_i})|}{\max \mathcal{P}}$, where $\max \mathcal{P}$ is as defined in Line 12 of Algorithm \ref{alg:filter_modified}. Thus, 
    \[
        \sum_{r \in R_i} (v_i \cdot(r - \mu_{X_i}))^2 = \sum_{x \in X_i} (v_i \cdot(x- \mu_{X_i}))^2 1(x \in R_i),
    \]
    where $1(\cdot)$ denotes the indicator random variable, and $\IE[1(x \in R_i)] = p_x$. By linearity of expectation, and the fact that conditional on a fixed $X_i$, $(v_i \cdot(x- \mu_{X_i}))^2$ is a constant, we have the following:
    \[
    \begin{split}
        \IE\left[\sum_{r \in R_i} (v_i \cdot(r - \mu_{X_i}))^2\right] &= \IE\left[\sum_{x \in X_i} (v_i \cdot(x- \mu_{X_i}))^2 1(x \in R_i)\right] \\
        &= \sum_{x \in X_i} (v_i \cdot(x- \mu_{X_i}))^2 \IE\left[1(x \in R_i)\right] \\
        & = \sum_{x \in X_i} (v_i \cdot(x- \mu_{X_i}))^2 p_x.        
    \end{split}
    \]
    By Chebyshev's sum inequality,
    \[
    \begin{split}
    &\frac{1}{|X_i|}\sum_{x \in X_i} (v_i \cdot(x- \mu_{X_i}))^2 p_x\\
    & \geq \left(\frac{1}{|X_i|}\sum_{x \in X_i} (v_i \cdot(x- \mu_{X_i}))^2\right)  \left(\frac{1}{|X_i|}\sum_{x \in X_i} p_x \right)\\  
        & = \lambda_i \frac{\IE[|R_i|]}{|X_i|}.
    \end{split}
    \]
    Multiplying both sides by $|X_i|$, we obtain 
    \[
    \sum_{x \in X_i} (v_i \cdot(x- \mu_{X_i}))^2 p_x \geq \lambda_i |R_i|.
    \]
\end{proof}

    \begin{lemma} [\emph{Decreasing Eigenvalue Lemma}]\label{lemma:lambda_decrease}
        For $i < \tau$, if $|v_i\cdot(\mu_{S_i'} - \mu_{X_i})|  > \delta'$, where $v_i$ is dominant eigenvector computed in iteration $i$ and $\delta'$ is as defined in Lemma \ref{lemma:expected_removal}, then $\IE[\lambda_{i}] > \IE[\lambda_{i+1}]$ holds on expectation (over the randomness in iteration $i$).
    \end{lemma}
    \begin{proof}
        Partition $X_i = G_i \cup R_i$, where $R_i$ are samples removed in iteration $i$ and $G_i = X_i \setminus R_i = X_{i+1}$. We will show that  if $\IE[\lambda_{i}] \leq \IE[\lambda_{i+1}]$, there exists a unit vector $v_i'$ along which the eigenvalue is larger than $\lambda_i$, contradicting that $\lambda_i$ is the largest possible.

        Let $v_{i+1}$ denote dominant eigenvector for $\Sigma_{X_{i+1}} = \Sigma_{G_{i}} $. Then, 
        \[
        \begin{split}
            &\sum_{g \in G_i} (v_{i+1}\cdot (g-\mu_{X_i}))^2\\
            &= \sum_{g \in G_i} (v_{i+1}\cdot ((g-\mu_{X_{i+1}}) + (\mu_{X_{i+1}} - \mu_{X_i})))^2 \\ 
            &= \sum_{g \in G_i} (v_{i+1}\cdot (g-\mu_{X_{i+1}}))^2 +|G_i|(v_{i+1}\cdot(\mu_{X_{i+1}} - \mu_{X_i}))^2 \\
            & \qquad + 2 v_{i+1}\cdot(\mu_{X_{i+1}} - \mu_{X_i}) \sum_{g \in G_i} v_{i+1}\cdot (g-\mu_{X_{i+1}})  \\
            & = |G_i| \lambda_{i+1} +  |G_i|(v_{i+1}\cdot(\mu_{X_{i+1}} - \mu_{X_i}))^2 \\
            & >|G_i| \lambda_{i+1} 
        \end{split}
        \]
        where the last equality is due to the fact that:
        $$ \sum_{g \in G_i} v_{i+1}\cdot (g-\mu_{X_{i+1}}) =  \sum_{x \in X_{i+1}} v_{i+1}\cdot (x-\mu_{X_{i+1}})=0$$  
        and $\frac{\sum_{g \in G_i} (v_{i+1}\cdot (g-\mu_{X_{i+1}}))^2}{|G_i|} = \displaystyle{\max_{||v_{i+1}||_2=1}}~(v_{i+1}\cdot \Sigma_{X_{i+1}}\cdot v_{i+1}) = \lambda_{i+1}$ 
        is the maximum eigenvalue of $\Sigma_{X_{i+1}} = \frac{\sum_{g \in X_{i+1}} (g-\mu_{X_{i+1}})^T(g-\mu_{X_{i+1}})}{|X_{i+1}|}$.

        Let $v_i'$ be a unit vector at angle $\theta_G$ and $\theta_R$ from $v_{i+1}$ and $v_i$, respectively.
        Then, 
        \[
        \begin{split}
            \sum_{g \in G_i} ((v_{i'}\cdot (g-\mu_{X_i}))^2 &= \sum_{g \in G_i} ((v_{i+1} \cos \theta_G \cdot (g-\mu_{X_i}))^2 \\
            &=\cos^2 \theta_G \sum_{g \in G_i} ((v_{i+1}\cdot (g-\mu_{X_i}))^2 \\
            & > \cos^2 \theta_G  |G_i| \lambda_{i+1} \\
            & > \cos^2 \theta_G  |G_i| \lambda_{i} \text{ (Assumption: $\lambda_{i+1} \geq \lambda_i$)}.
        \end{split}
        \]
        
       
        Similarly, denoting $z_i = v_i \cdot(r-\mu_{X_i})$, we get:
        \[
            \sum_{r \in R_i} (v_i' \cdot(r-\mu_{X_i}))^2 = \cos^2 \theta_R \sum_{i \in R_i} (z_i^2) \geq \cos^2 \theta_R |R_i|\lambda_i
        \]

        where the last inequality uses Lemma \ref{lemma:helper}, which shows that that for points removed $\IE[\sum_{i \in R_i} (z_i^2)] > \lambda_i\IE[|R_i|]$ . Thus,

        \[
        \begin{split}
            \sum_{x \in X_i} (v_i' \cdot (x - \mu_{X_i}))^2 &=\sum_{g \in G_i} ((v_{i'}\cdot (g-\mu_{X_i}))^2  + \sum_{r \in R_i} (v_i' \cdot(r-\mu_X))^2 \\
            &> \cos^2 \theta_G  |G_i| \lambda_{i} + \cos^2 \theta_R |R_i| \lambda_i \\
            & = (1-\sin^2 \theta_G ) |G_i| \lambda_{i} + (1-\sin^2 \theta_R) |R_i| \lambda_i \\
            &= \lambda_i (|R_i| + |G_i|) - (\sin^2 \theta_G  + \sin^2 \theta_R) \\
            & \geq \lambda_i (|R_i| + |G_i|-1)
        \end{split}
        \]
        We used the following geometric fact: There always exists a $\theta_G$, such that $(\sin^2 \theta_G  + \sin^2 \theta_R) \leq 1$.
Now, 
    \[
        \lambda_i' = \frac{1}{|R_i|+|G_i|} \sum_{x \in X_i} (v_i' \cdot (x - \mu_{X_i}))^2  > \lambda_i \left(1 - \frac{1}{|R_i|+|G_i|} \right) \approx \lambda_i
    \]
    since $1/(|R_i|+|G_i|) = \Omega(1/n)$ which is negligible (within $\stackrel{f}{=}$ check used for eigenvalue convergence check for large $n$).
    \end{proof}

The following lemma summarizes the loop invariant invariant in Lemma \ref{lemma:loop_invariant} for the main loop in Algorithm~\ref{alg:filter_modified}.  

\begin{lemma} \label{lemma:loop_invariant}
    Algorithm \ref{alg:filter_modified} admits the following loop invariant: At the start of each iteration $i \leq \tau$, all the following conditions hold:
    \begin{enumerate}
        \item $|X_i| \geq (1-5\epsilon)n$ with probability $1- \exp(-\Omega(n\epsilon))$.
        \item  $\lambda_i \leq \lambda_{i-1}$, with  $\lambda_0=\infty$,  on expectation. 
        \item $i \leq 2n\epsilon$.
    \end{enumerate}
    if $|v_i\cdot(\mu_{S_i'} - \mu_{X_i})|  > \delta'$, where $v_i$ is the dominant eigenvector in iteration $i$, and $|T'_i|>0$.
\end{lemma}
\begin{proof}
    For the case of $i=1$, all conditions of the invariant are trivially satisfied. Thus, we consider the case where $i \in [2, \tau]$.  

    For Condition (1), we appeal to Corollary \ref{corr:whp_difference}. By Corollary \ref{corr:whp_difference}, $\IP[N_S - N_T \geq n\epsilon] \leq \exp(-\Omega(n\epsilon))$, where $N_S$ and $N_T$ are the number of samples removed from $S'$ and $T'$ after $\tau$ iterations, respectively. Since there are at most $n\epsilon$ naturally occurring outliers (Corollary \ref{corr:max_deviation}) and $n\epsilon$ corrupted samples,  $N_T \leq |T_1| \leq  2n\epsilon$. Thus, $N_S \leq 2n\epsilon + n\epsilon = 3n\epsilon$, or equivalently,  $|S_{i}| \geq |S_1| - N_S \geq (1-2\epsilon)n - 3\epsilon = (1-5\epsilon)n$ with probability at least $1-\epsilon(-\Omega(n \epsilon))$ across all $i$.

Condition (2) is direct consequence of Lemma~\ref{lemma:lambda_decrease} with randomness used in iteration $i-1$ . 

    For Condition (3), since the algorithm has at most $2n\epsilon$ iterations, this condition is trivially satisfied.  
\end{proof}

Finally, Theorem \ref{thm:bias_revised} proves the correctness of Algorithm \ref{alg:filter_modified}, by showing that on expectation, if the algorithm terminates at any one of the three stopping conditions, $\IE[|v\cdot(\mu_{X_{\tau+1}} - \mu_S)|] \leq \delta'$ which is $O(\delta)$ for fixed $\epsilon \leq \frac{1}{12}$, This error bound is the same bound derived the original filtering algorithm (Algorithm \ref{alg:filter_original})\footnote{Given in \cite{diakonikolas2023algorithmic}, Theorem 2.17}.
\begin{theorem} \label{theorem:max_bias}
    Let $S$ be a $(5\epsilon, \delta)$-stable set, where $\epsilon \leq 1/12$, 
    $\delta = \sqrt{20 ||\Sigma_S||_2}$, and $X$ is constructed by corrupting up to an $\epsilon$ proportion of samples from $S$. 
    Suppose Algorithm \ref{alg:filter_modified} terminates at $\tau \leq 2n\epsilon$
     number of iterations. Let $X_{\tau+1}$ be the samples remaining after $\tau$ iterations. Then, for any unit vector $v$, $|v \cdot (\mu_{X_{\tau+1}} - \mu_S)| \leq \delta'$, where $\delta' = 2\frac{\delta}{\epsilon}+\delta$. 
\end{theorem}
\begin{proof} 
    For $i\leq \tau$, let  $X_i$ denote the remaining points in $X$ at the start of the $i$-th iteration, and $\epsilon'_i$ denote
    the proportion of samples remaining in outlier set $T_i'$, where $T_i'$ is as defined in Definition \ref{definition:outlier}. 
    
    We have established in Lemma~\ref{lemma:loop_invariant} at the start of each iteration $i \in [1,\tau]$, the loop invariant conditions hold if the following holds: 
    
    $$(|T'_i| >0) \land (|v_i\cdot(\mu_{S_i'} - \mu_{X_i})|  > \delta')$$
    
    where $v_i$ is the dominant eigenvector for $\Sigma_{X_i}$. We call this the continuation condition. We will now show that the continuation condition is false at $i=\tau+1$, i.e., at the end of last iteration $\tau$. 
    
    Consider each of the $3$ loop exit conditions:
    \begin{enumerate}
        \item $\tau = 2n\epsilon$: By Lemma \ref{lemma:expected_removal}, either $|T_i| = 0$ or we remove at least 1 sample from $T_i$ on expectation in every iteration $i$. Since by Corollary \ref{corr:max_deviation}, $|T_1| \leq 2n\epsilon$, we have $|T_{\tau+1}|=|T_{2n\epsilon}| = 0$.
        
        \item $|X_{\tau+1}| \leq (1-4\epsilon)n$ (Line 9): For exit condition to be fulfilled, the number of inliers removed $N_S$ and outliers removed $N_T$ should satisfy $N_T + N_S > 4\epsilon n$. But since $N_T$ is at most $2n\epsilon$, we must have $N_S > 2\epsilon n$. Corollary \ref{corr:whp_difference} shows that we have $N_S - N_T > \epsilon n$ with probability at most $\exp(-\Omega(n\epsilon))$ as long as $T_{\tau}$ has at least one point in it. Hence, when this condition triggers at some $\tau$, $|T'_{\tau}|=0=|T'_{\tau+1}|$, and $|S'_{\tau+1}| \geq (1-5\epsilon)n$ with probability at least  $1-\exp(-\Omega(n\epsilon))$.
        
        
        \item Eigenvalue convergence check (Line 3) is fulfilled: By the decreasing eigenvalue lemma (Lemma \ref{lemma:lambda_decrease}) shows that when the continuation condition holds $\IE[\lambda_{i+1}] < \IE[\lambda_{i}]$. Therefore, in iteration $i=\tau$, $\IE[\lambda_{\tau}] < \IE[\lambda_{\tau-1}]$. Since the convergence check is fulfilled, $\lambda_\tau = \lambda_{\tau-1}$, which contradicts the expected inequality. Therefore, the continuation condition is false on expectation.

        
    \end{enumerate}    

        We have shown that in each of the 3 cases, the continuation condition is false on expectation. If $|T'_{\tau+1}|=0$, $X_{\tau+1} = S'_{\tau+1}$. By Definition of stability we have $|v\cdot(\mu_{S'_{\tau+1}} - \mu_{X_{\tau+1}})| \leq \delta'$ for all unit vectors $v$. This happens with probability at least $1-\exp(-\Omega(n\epsilon))$ in the case (1) or (2) above. In case (3), if $|T'_{\tau+1}|\neq 0$, we have that continuation condition is false because 
        $|v_i\cdot(\mu_{S'_{\tau+1}} - \mu_{X_{\tau+1}})| \leq \delta'$.
        Since $v_i$ is the dominant eigenvector and the $X_{\tau}$ is unchanged in $\tau$, we have that $|v\cdot(\mu_{S'_{\tau+1}} - \mu_{X_{\tau+1}})| \leq \delta'$ for all unit vectors $v$. Thus, in all cases, $\forall{v},|v\cdot(\mu_{S'_{\tau+1}} - \mu_{X_{\tau+1}})| \leq \delta'$ on expectation.
    \end{proof}

    Now, we can show that bias is $O(\sqrt{\epsilon})\cdot||\Sigma_S||_2$.

    \begin{theorem*}[\emph{Theorem \ref{thm:bias_revised} restated}] 
    Let $S$ be a $(5\epsilon, \delta)$-stable set, where $\epsilon \leq 1/12$, 
    $\delta = \sqrt{20} \sqrt{||\Sigma_S||_2}$, and $X$ is constructed by corrupting an $\epsilon$ proportion of $S$. 
    Suppose the main loop in Algorithm \ref{alg:filter_modified} terminates after $\tau \leq 2n\epsilon$ iterations with $X_{\tau+1}$  as samples left at Line 15.  Then, on expectation,
        $||\mu_{X_{\tau+1}} - \mu_{S}||_2 \leq \beta \sqrt{||\Sigma_S||_2}$, for $\beta = \sqrt{20} \left(\frac{2}{\epsilon} + 2\right)$. 
    \end{theorem*}
    \begin{proof}
        From Theorem \ref{theorem:max_bias}, for all unit vector $v$, $|v \cdot (\mu_{X_{\tau+1}} - \mu_{S'_{\tau+1}})| \leq \delta'$, where $\delta' = 2\frac{\delta}{\epsilon}+\delta$. 
        Thus, 
        \[
            ||\mu_{X_{\tau+1}} - \mu_{S'_{\tau+1}}||_2 = |v \cdot (\mu_{X_{\tau+1}} - \mu_{S'_{\tau+1}})| \leq \delta' 
        \]
        and since Lemma~\ref{lemma:loop_invariant} shows that $|S_{\tau+1}| \geq (1-5\epsilon)n$ with all but negligible probability, by Definition~\ref{dfn:epsilon_delta_stable}:
    \[
        \frac{\left| \displaystyle{\sum_{s \in S'_{\tau+1}}} v\cdot(s-\mu_S)\right|}{|S'_{\tau+1}|}   \leq \frac{\displaystyle{\sum_{s \in S'_{\tau+1}}} |  v\cdot(s-\mu_S) |}{|S'_{\tau+1}|}  \leq ||\mu_{S'_{\tau+1}} - \mu_S||_2 < \delta
    \]        
        By then using triangle inequality,
        \[
            ||\mu_{X_{\tau+1}} - \mu_S||_2 \leq \delta' + \delta = \delta (\frac{2}{\epsilon} + 2)
        \]
        Since $\delta = \sqrt{20 ||\Sigma_S||_2}$, the claim follows for $\beta = \sqrt{20} \left(\frac{2}{\epsilon} + 2\right)$
    \end{proof}

For a fixed constant $\epsilon$, the bias of Algorithm~\ref{alg:filter_modified} is $O(1)\sqrt{||\Sigma_S||_2}$.   

\section{Code Implementation}
The code implementation of \randeigen~ for both the federated and centralized setup is provided in the following GitHub repository: \url{https://github.com/dezhanglee/randeigen_artifacts/tree/master}. This implementation builds on the implementations provided by \cite{zhu2023byzantine} and \cite{DBLP:conf/sp/ChoudharyKS24}.
\section{Aggregation on Other Optimizers} \label{appendix:adam}

\randeigen can be integrated into steps involving client data aggregation in the federated learning setting. We illustrate this in Algorithm \ref{alg:fedavg_adam}, where
we adapt an existing federated learning implementation which employs the Adam optimizer \cite{DBLP:conf/iclr/ReddiCZGRKKM21}, which incorporates \randeigen~ in Line 
\ref{alg:fedavg_adam_line_agg}. 

\begin{algorithm}[H]
\caption{Federated Averaging (FedAvg) \cite{DBLP:conf/aistats/McMahanMRHA17}
using Adam~\cite{adam}}
\label{alg:fedavg_adam}
\begin{algorithmic}[1]
\STATE \textbf{Input:} Number of training steps $T$, clients per step $K$, learning rate $\eta$, decay paramters $\beta_0, \beta_1 \in [0,1)$, adaptivity parameter $\tau$,
initial model parameters $w_0$

\FOR{step $t = 0, 1, \dots, T-1$}
    \STATE Server sends $w_t$ to all $k$ clients
    \FOR{each client $k \in \mathcal{S}_t$ \textbf{in parallel}}
        \STATE Client $k$ trains model with parameters $w_t$ locally
        \STATE Compute  gradient of loss function $\nabla \mathcal{L}_k(w_t)$
        \STATE Send gradients $\nabla \mathcal{L}_k(w_t)$ back to the server.
    \ENDFOR
    
    \STATE  $\nabla \mathcal{L}_{t} = \randeigen(\nabla \mathcal{L}_1(w_t), \dots, \nabla \mathcal{L}_k(w_t))$ \label{alg:fedavg_adam_line_agg}
    \IF{$t > 0$}
        \STATE $m_t = \beta_1 m_{t-1} + (1-\beta_1)\nabla \mathcal{L}_{t}$
        \STATE $v_t = \beta_2 v_{t-1} + (1-\beta_2) (\nabla \mathcal{L}_{t})^2$
    \ELSE 
        \STATE $m_t = \nabla \mathcal{L}_{t}$
        \STATE $v_t = (\nabla \mathcal{L}_{t})^2$
    \ENDIF
    \STATE Server computes $w_{t+1} = w_t - \eta ( m_t/(\tau+\sqrt{v_t}))$
\ENDFOR

\STATE \textbf{Output:} Final global model with parameters $w_T$.
\end{algorithmic}
\end{algorithm}

\section{Deferred Proofs} \label{appendix:proof}

\begin{lemma} \label{lemma:moments_A}
    Let $A$ be a $d \times k$ matrix, where each entry in $A$ is an i.i.d. $\mathcal{N}(0,1)$ random variable. Then, $\IE[AA^T] = kI$, $Cov(A^T, AA^T) = 0$ and thus, $\IE[A^T A A^T] = \IE[A^T] \IE[AA^T]$.
\end{lemma}

\begin{proof}
    To derive $Cov(A^T, AA^T) = \IE[A^T A A^T] - \IE[A^T] \IE[ A A^T]  $, we first need to determine $\IE[A^T]$, $\IE[ A A^T]$, and $\IE[A^T A A^T]$. These are the first, second and third moments of the random matrix $A$. 

    Let $(A)_{ij}$ denote the $(i,j)$-th entry of $A$. Then, for all $i, j$, it is immediate from the construction of $A$ that $\IE[(A)_{ij}] = 0$ and $\IE[A] = 0I$, where $I$ is the identity matrix. 
    
    Next, we derive the second moment of $A$. For all $1 \leq i \leq d$, $(AA^T)_{ii}$ is a $\chi^2_k$ random variable, and hence $\IE[(AA^T)_{ii}] = k$. On the other hand, for all $i \neq j$, $(AA^T)_{ij}$ is the dot product of two independent multivariate Gaussian vectors with mean 0, and hence, $\IE[(AA^T)_{ij}] = 0$. Thus,  $\IE[AA^T] = k I$.

    To derive the third moment of $A$, we rely on the following results in probability theory. 
    \begin{enumerate}
        \item If $m(t)$ is the moment generating function of $A$, then $h$-th moment of $A$ is $ \IE[A^h] = m^{(h)}(0)$, where $m^{(h)}(t)$ denotes the $h$-th derivative of $m(t)$ (\cite{gut2013probability}, Theorem 8.3 (ii)).
        \item $A$ follows a matrix normally distributed random variable. This can be written as $A\sim N_{d, k} (\mu, \Sigma, \Psi)$, where $\mu=0$ denotes the row-wise mean, and $\Sigma = I_d$ and $\Psi = I_k$ denotes the row-wise and column-wise covariance matrices \cite{von1988moments}.
        \item It can be shown (\cite{von1988moments}, Section 3) that the moment generating function admits a closed-form expression
        \[
            m(t) = \exp\left( \frac{1}{2}Trace(t^T \Sigma t \Psi) \right).
        \]
        \item It can also be shown (\cite{von1988moments}, proof of Theorem 3.1) that $A$ admits the  third moment $\IE[A^T A A^T]=m'''(0) = 0$.
    \end{enumerate}
    Therefore, $Cov(A^T, AA^T) = \IE[A^T A A^T] - \IE[A^T] \IE[A^T A A^T] = 0 - 0 (kI) =0 $. Thus, from the definition of covariance, it follows that $\IE[A^T A A^T] = \IE[A^T] \IE[A A^T]$
\end{proof}






\begin{proof}[\emph{{Proof of Theorem \ref{thm:rs_ujl_u} }}]
    Let $\lambda_1, \dots, \lambda_d$ denote the eigenvalues
    of $\X$, where $|\lambda_1| \geq  \dots \geq |\lambda_d|$, and let $u_i$ be the eigenvector associated with
    $\lambda_i$. 
Let $v = (1/\sqrt{k}) A^T u$ , where $u:=u_1$ is the dominant eigenvector of $\X^T \X$. 

We will first show that $v$ is an eigenvector of $\XJL^T \XJL$. Observe that
\[
\begin{split}
    \XJL^T \XJL v = A^T \X^T \X A v = (1/\sqrt{k}) A^T \X^T \X A  A^T u.
\end{split}
\]
Taking expectation over the choices of $A$,
\[
\begin{split}
   \IE[Y^T Y v ]& = \IE[(1/\sqrt{k}) A^T X^T X A  A^T u]\\
   & \stackrel{1}{=} \IE[(1/\sqrt{k}) A^T X^T X] (kI) u\\
    & =   \IE[\sqrt{k} A^T X^T X u]\\
    &= \IE[\sqrt{k} A^T \lambda_1 u] \\
    &  \stackrel{2}{=} \IE[ \lambda_1 k (1/\sqrt{k}) A^T u] \\
    &= k \lambda_1\IE[ v],
\end{split}
\]
    where  $\stackrel{1}{=}$ is a consequence of Lemma \ref{lemma:moments_A} that (a) $\IE[AA^T] = kI$, and (b) $A^T$ and $AA^T$ are uncorrelated and the expectation can thus be separated. In $\stackrel{2}{=}$, we used the fact that since $u$ is the dominant eigenvector of $\X^T \X$, $\X^T \X u = \lambda_1 u$. Therefore, on expectation, $v$ is an eigenvector of $\XJL^T \XJL$ with eigenvalue $k \lambda_1$.
    Furthermore, it can also be verified that $\IE[||(1/\sqrt{k}) A^T u||_2] = ||u||_2 = 1$  (i.e. on expectation, $v$ is a vector of unit-norm). This follows from the fact that each coordinate of $A^Tu$ is $\mathcal{N}(0,||u||_2)$ and thus $\IE[||A^Tu||_2^2]=k$. 
    
    Next, we show that  $\hat{\lambda}_1:= k\lambda_1$ is the dominant eigenvalue of $\XJL^T \XJL$. Suppose $\lambda^*_1$ is another eigenvalue
    of $\XJL^T \XJL$ with a corresponding eigenvector $v'$, 
    where $|\lambda^*| > |\hat{\lambda}_1|$. Then, the following holds:
    \[
        \lambda^*_1 v' = \XJL^T \XJL v' = \frac{1}{k}A^T \X^T \X A v'.
    \]
    Multiplying both sides by $A$ and considering the leftmost and rightmost terms, 
    \[
         \lambda^*_1 A v' = \frac{1}{k}AA^T \X^T \X A v'.
    \]
    Taking expectations and using the fact that $\IE[AA^T \X^T \X A v'] = \IE[AA^T] \IE[ \X^T \X A v']$ (Lemma \ref{lemma:moments_A}), 
    \[
        \lambda^*_1 \IE[ A v']  = \frac{1}{k} \IE\left[ AA^T\right] \IE[\X^T \X A v'] = \X^T \X \IE[A v'], 
    \]
    where we used $\IE[AA^T] = kI$ (Lemma \ref{lemma:moments_A}) in the last equality. This implies that $\IE[ A v'] $ is the dominant eigenvector of $\X^T \X $ associated with a dominant eigenvalue $|\lambda^*_1| > |\lambda_1|$. This is a contradiction since $\lambda_1$ is the dominant eigenvalue of $\X^T\X$. Thus, we have shown that on expectation, $A v'$ is the dominant eigenvector of $\XJL^T \XJL$.  
\end{proof}

In our proof of Theorem \ref{thm:dist_jl}, we will use Corollary 2.1 from \cite{RandomProjectionsLecture}, which is a corollary of the Johnson–Lindenstrauss lemma. For convenience, this result is restated in Corollary \ref{corr:jl_proj}.
\begin{corollary}[\emph{Corollary 2.1 from \cite{RandomProjectionsLecture}}] \label{corr:jl_proj}
    Let $a, b \in \mathbb{R}^d$, and that $||a||_2 \leq 1$ and $||b||_2 \leq 1$. Let $f(x) = \frac{1}{\sqrt{k}} x A$, where $A$ is a $d \times k$ matrix, and each entry is sampled i.i.d from $\mathcal{N}(0,1)$. Then, 
    \[
    \IP[ | a \cdot b - f(a) f(b) | \geq \epsilon_{JL}] \leq 4\exp\left(- \frac{(\epsilon_{JL}^2 - \epsilon_{JL}^3)k}{4} \right)
    \]
\end{corollary}

\begin{proof} [\emph{Proof of Theorem \ref{thm:dist_jl}}]
    Given $x = (x_1, \dots, x_d)$ and $u = (u_1,\dots, u_d)$, let $\hat{x} = (x_1 u_1, \dots, x_d u_d)$ denote the components of $x$ after projection on the unit vector $u$. Let $P_X =\max \mathcal{P}$ be as defined in Algorithm \ref{alg:filter_modified}. By definition of $P_X$ (which is the sample $x \in X$ which has the largest projection on $u$), $\left| \left| \frac{\hat{x}}{P_X} \right| \right|_2 \leq 1$. Thus, by Corollary \ref{corr:jl_proj}, the following holds:
    \[
    \IP\left[ \left| \frac{\hat{x}}{P_X} \cdot u - f\left(\frac{\hat{x}}{P_X}\right) f(u) \right| \geq \epsilon_{JL}\right] \leq 4\exp\left(- \frac{(\epsilon_{JL}^2 - \epsilon_{JL}^3)k}{4} \right).
    \]
    Using $\epsilon_{JL} = 0.1$ and $k = \frac{\log d}{\epsilon_{JL}^2}$ (which are the values used in Algorithm \ref{alg:randeigen}),
    \begin{equation} \label{eqn:conc_alg_randeigen}
    \begin{split}
    \IP\left[ \left| \frac{\hat{x}}{P_X} \cdot u - f\left(\frac{\hat{x}}{P_X}\right) f(u) \right| \geq \epsilon_{JL}\right] & \leq 4\exp\left(- \frac{(0.1^2 - 0.1^3)\frac{\log d}{0.1^2}}{4} \right) \\
    &= \frac{4}{d^{0.225}}.
    \end{split}
    \end{equation}
\end{proof}
\begin{proof}[\emph{Proof of Corollary \ref{corr:prob_diff}}]
     By Corollary \ref{corr:jl_proj},
    \begin{equation}       \label{eqn:conc_px_py}  
    \begin{split}
        &\IP\left[\left|\frac{\hat{x}}{||\hat{x}||_2} \cdot u - f\left(\frac{\hat{x}}{||\hat{x}||_2}\right) \cdot f(u) \right| > \epsilon_{JL}\right] \\
        & = \IP\left[\left|\hat{x}\cdot u - f\left(\hat{x}\right) \cdot f(u) \right| > \epsilon_{JL} ||\hat{x}||_2\right] \\
        & = \IP\left[\left|P_X - f\left(\hat{x}\right) \cdot f(u) \right| > \epsilon_{JL} ||\hat{x}||_2\right] \\
        & \leq 4\exp\left(- \frac{(\epsilon_{JL}^2 - \epsilon_{JL}^3)k}{4} \right) = \frac{4}{d^{0.225}}.
    \end{split}
    \end{equation}
    Let $P_Y = \max \mathcal{P}$ as defined in Algorithm \ref{alg:randeigen}, $P_X$ be as defined in Theorem \ref{thm:dist_jl}, and  
    \begin{equation} \label{eqn:x'}
        x' = \arg \max_{x \in X} |u \cdot (x -\mu_X)|, 
    \end{equation}
    which is the sample in $X$ used to compute $P_X$. Since $\hat{x}$ is a projection of a given $x \in X$ on $u$, it follows that $||\hat{x}||_2 \leq ||\hat{x}'||_2= P_X$. Thus, we can bound (\ref{eqn:conc_px_py}) as,
    \begin{equation}       \label{eqn:conc_px_py_2}  
    \begin{split}
       &  \IP\left[\left|P_X - f\left(\hat{x}\right) \cdot f(u) \right| > \epsilon_{JL} P_X \right]\\
       & \leq \IP\left[\left|P_X - f\left(\hat{x}\right) \cdot f(u) \right| > \epsilon_{JL} ||\hat{x}||_2\right]     \leq  \frac{4}{d^{0.225}}.
    \end{split}
    \end{equation}
    Consider $y' = \arg \max_{y \in Y} |v \cdot (y - \mu_Y)|$, where $v$ is the dominant eigenvector of $\Sigma_Y$ and let $x''$ be the sample used to compute $y'$ (i.e. $y' = f(x'')$). From Theorem \ref{thm:rs_ujl_u}, $f(u)$ is the expected dominant eigenvector of $\Sigma_Y$. Henceforth, we will consider projections on $f(u)$. Thus, 
    \[
        \IE[P_Y] = \IE[|v \cdot (y' - \mu_Y)|] = \IE \left[ \frac{1}{k}\left| u A A^T (x'' - \mu_X)\right| \right] = |u \cdot(x'' - \mu_X)|,
    \]
    which is maximized when $x'' = x'$ (Due to the definition of $x'$ in \ref{eqn:x'}).  
    When samples in $Y$ are projected on $v$, $y' = f(x')$. Hence,  $\IE[P_Y] = \IE \left[ f(\hat{x}') \cdot f(u) \right]= \IE \left[\frac{1}{k} \hat{x}'A A^T u\right] = \hat{x}' u$. Henceforth, we will let $P_Y = f(\hat{x}') \cdot f(u)$ since we consider projections of each $y \in Y$ on the expected dominant eigenvector $f(u)$.
    From (\ref{eqn:conc_px_py_2}),
    \[
    \begin{split}
    &\IP[|P_X - f(\hat{x}') \cdot f(u)| > \epsilon_{JL}P_X]\\
    &=\IP[|P_X - P_Y| > \epsilon_{JL}P_X]\\
        &= \IP\left[P_Y \not\in (P_X(1 -\epsilon_{JL}), P_X(1+\epsilon_{JL}) )\right] \\
        &= \IP\left[\frac{P_X}{P_Y} \not \in \left(\frac{1}{1+\epsilon_{JL}}, \frac{1}{1-\epsilon_{JL}} \right)\right]\leq \frac{4}{d^{0.225}}.
    \end{split}
    \]
    Taking a union bound with (\ref{eqn:conc_alg_randeigen}), 
    \[
    \begin{split}
       &  \IP \Bigg[(\left|\hat{x}\cdot u - f\left(\hat{x}\right) \cdot f(u) \right| > \epsilon_{JL} P_X) \\
       & \qquad \qquad \cup \left( \frac{P_X}{P_Y}\not\in \left(\frac{1}{1+\epsilon_{JL}}, \frac{1}{1-\epsilon_{JL}} \right) \right) \Bigg]  \\
       &\leq  \frac{4}{d^{0.225}} +  \frac{4}{d^{0.225}}=  \frac{8}{d^{0.225}}.
    \end{split}
    \]
    Hence, with a probability of at least $1-\frac{8}{d^{0.225}}$,
    \[
        \frac{P_X}{P_Y}\epsilon_{JL} \in \left(\frac{\epsilon_{JL}}{1+\epsilon_{JL}}, \frac{\epsilon_{JL}}{1-\epsilon_{JL}} \right)
    \]
    and therefore, 
    \[
        \frac{1}{P_Y}\left|\hat{x}'\cdot u - f\left(\hat{x}'\right) \cdot f(u) \right| \leq  \frac{P_X}{P_Y}\epsilon_{JL} \leq \frac{\epsilon_{JL}}{1-\epsilon_{JL}} = \frac{1}{9}.
    \]
    Since 
    \[
    \frac{1}{P_Y}\left(\hat{x}\cdot u - f\left(\hat{x}\right) \cdot f(u) \right)\leq\frac{1}{P_Y}\left|\hat{x}\cdot u - f\left(\hat{x}\right) \cdot f(u) \right|,
    \]
    and by linearity of expectation, 
    \[
    \begin{split}
        \IE\left[ \frac{1}{P_Y}\left(\hat{x}\cdot u - f\left(\hat{x}\right) \cdot f(u) \right)\right] &= \IE\left[\frac{\hat{x}\cdot u }{P_Y} - \frac{f\left(\hat{x}\right) \cdot f(u) }{P_Y}\right] \\
        &=\IE\left[\frac{\hat{x}\cdot u }{P_Y} \right] - \IE \left[\frac{f\left(\hat{x}\right) \cdot f(u) }{P_Y}\right] \\
        & = \frac{\hat{x}\cdot u }{\IE[P_Y]} - p_{y}\\
        & = \frac{\hat{x}\cdot u }{\IE[\frac{1}{k} x' A A^T u]} - p_{y} \\
        & =  p_{x} - p_{y}.
    \end{split}
    \]
    So, with probability at least $1-\frac{8}{d^{0.225}}$, $|p_{x} - p_{y}| \leq \frac{\epsilon_{JL}}{1-\epsilon_{JL}}$.
\end{proof}
\begin{proof}[\emph{Proof of Theorem \ref{thm:power_iteration_error}}]
    For the first claim on $k$, \cite{Gharan2017} showed that
    \[
        \frac{v_N M v_N^T}{v_Nv_N^T} \geq (1-\epsilon_p) \lambda_1 \frac{1}{1+4k(1-\epsilon_p)^{2N}},
    \]
    which decreases with $N$.
    Consider 
    $N \geq -\frac{\log(4k)}{log (1-\epsilon_p)} = -\log_{1-\epsilon_p} (4k)$. Then, 
    \[
    \begin{split}
        (1-\epsilon_p) \lambda_1 \frac{1}{1+4k(1-\epsilon_p)^{2N}} & \geq  (1-\epsilon_p) \lambda_1 \frac{1}{1+4k\frac{1}{(4k)^2}} \\
        &=    (1-\epsilon_p) \lambda_1 \frac{1}{1+\frac{1}{4k}}.
    \end{split}
    \]
    which approaches $(1-\epsilon_p) \lambda_1 $ as the size of the matrix $k$ increases. In Algorithm \ref{alg:randeigen}, $k = 100\log d$, and thus the above lower bound becomes $(1-\epsilon_p) \lambda_1 \frac{1}{1+\frac{1}{400 \log d}}$.
    
    Next, we show that
    $||v_N - u_1||_2$ is decreasing in $N$. Let $v_0$ be the randomly generated
    initial starting point. Since $M$ is of full rank and spans $\mathbb{R}^k$, there exists unit vectors $c_1, \dots, c_k $ such that
    $v_0 = c_1 u_1 + \dots + c_k u_k$,
    and hence,
    \[
    \begin{split}
        v_N = A^N v_0 & = \lambda_1^N \left(c_1 u_1 + \sum_{i=2}^k c_i u_i \left( \frac{\lambda_i}{\lambda_1}\right)^N \right) \\
        & \leq \lambda_1^N \left(c_1 u_1 + \left( \frac{\lambda_2}{\lambda_1}\right)^N \sum_{i=2}^k c_i u_i  \right) 
    \end{split}
    \]
    The upper bound, after $\ell_2$ normalization, is therefore,
    \[
        \frac{v_N}{||v_N||_2} \leq u_1 + O\left( \left(\frac{\lambda_2}{\lambda_1}\right)^{2N} \right) \sum_{i=2}^k c_i u_i.
    \]
    Hence, 
    \[
        \left| \left| \frac{v_N}{||v_N||_2} - u_1 \right| \right|_2 \leq  O\left( \left(\frac{\lambda_2}{\lambda_1}\right)^{2N} \right) \sum_{i=2}^k c_i u_i \in O\left( \left(\frac{\lambda_2}{\lambda_1}\right)^{2N} \right) .
    \]
    Set $N \geq \frac{\log \epsilon_p}{2 \log (\lambda_2 / \lambda_1)}$, then $\left(\frac{\lambda_2}{\lambda_1}\right)^{2N} \leq \epsilon_p$. Thus, it suffices to set  $N \geq  \max \left(\left| \frac{\log (4k)}{\log (1-\epsilon_p)} \right|, \left|  \frac{\log \epsilon_p}{2 \log (\lambda_2 / \lambda_1)}\right|\right)$.
\end{proof}


\end{document}